\documentclass[journal]{IEEEtran}
\usepackage[T1]{fontenc}

\usepackage{cite}

\usepackage[pdftex]{graphicx}

\ifCLASSINFOpdf
\else
\fi
\usepackage{amsmath}
\usepackage{amssymb}
\usepackage{color}

\newcommand{\NPE}{\ensuremath{N_{\text{PE}}}}
\newcommand{\FCLK}{\ensuremath{f_{\text{clk}}}}
\begin{document}
%
\title{A Distributed Processing Architecture for Modular and Scalable Massive MIMO Base Stations}
%
%
%


\author{
	Erik~Bertilsson,~\IEEEmembership{Student Member, IEEE},
	Oscar~Gustafsson,~\IEEEmembership{Senior Member, IEEE}, \\
	and~Erik~G.~Larsson,~\IEEEmembership{Fellow, IEEE}%
	\thanks{The authors are with the Department of Electrical Engineering, Link\"oping University, SE-581 83 Link\"oping, Sweden. Emails: \{\mbox{erik.bertilsson}, \mbox{oscar.gustafsson}, \mbox{erik.g.larsson}\}@liu.se}%
\thanks{Manuscript received \today.}
}

\maketitle

\begin{abstract}
	In this work, a scalable and modular architecture for massive MIMO base stations with distributed processing is proposed. New antennas can readily be added by adding a new node as each node handles all the additional involved processing. The architecture supports conjugate beamforming, zero-forcing, and MMSE, where for the two latter cases a central matrix inversion is required. The impact of the time required for this matrix inversion is carefully analyzed along with a generic frame format. 
	As part of the contribution, careful computational, memory, and communication analyses are presented. It is shown that all computations can be mapped to a single computational structure and that a processing node consisting of a single such processing element can handle a broad range of bandwidths and number of terminals. 
\end{abstract}


%
\IEEEpeerreviewmaketitle

\section{Introduction}\label{sec:introduction}
\IEEEPARstart{T}{he} ever increasing demands for higher data rates in wireless communication opens up for many opportunities and challenges in the fifth generation (5G) wireless infrastructure \cite{Boccardi2014,Agyapong2014}. One such is the use of many antennas on the base station side, commonly referred to as Massive MIMO or Very Large Scale MIMO \cite{Hoydis2013,Lu2014,Larsson2014,Bjoernson2016,marzetta2016fundamentals}. By using many antennas, compared to the number of terminals, the transmit and receive power of each antenna and the processing associated per antenna can be reduced. Furthermore, the total energy per bit per user can potentially be reduced at the system level compared to traditional few antenna solutions.


However, while massive MIMO is a promising technology there are still obstacles to overcome before systems of this type can be deployed. There exists a number of demonstrators \cite{shepard2012argos,Vieira2014,Yang2016,Luo2016} that have been used to demonstrate the feasibility of the techniques. The number of antennas for the demonstrators are typically between 64 and 128 and supports up to 12 terminals. In addition, some work has been done to implement parts or all of the involved processing \cite{Wu2014,Prabhu2015a,Puglielli2016,Prabhu2017,kaipeng2016distributed,li2016decentralized}, either using a centralized node with all processing \cite{Wu2014,Prabhu2015a,Prabhu2017} or distributing the processing to several nodes \cite{Puglielli2016,kaipeng2016distributed,li2016decentralized}. For the centralized node architectures, typically the case of 128 antennas and 8 terminals are considered.

Of more interest in our current context are the examples of distributed processing. In \cite{Puglielli2016}, a base station system design that is constructed by identical modules is proposed. The baseband processing is distributed among the modules, which are connected in an array. The modules contain RF-front ends, digital baseband processing and digital interconnection link to all four neighbors. The system design, along with cost and power consumption issues are analyzed. However, there are no details on how the baseband processing should be performed or the impact of timing constraints.

In \cite{kaipeng2016distributed, li2016decentralized}, the authors propose distributed processing based on COTS processors. However, the timing constraints of the considered LTE-frame structure is not taken into consideration. Additionally, the systems are not dimensioned to meet the maximum obtainable throughputs of the considered specifications.

Compared to a centralized architecture, in a distributed architecture, the number of antennas at the base station can more easily be scaled. For the centralized architecture, increasing the number of antennas more or less requires a complete redesign of the system. In a distributed architecture, the number of antennas can be increased by adding another node that contains the antenna and circuitry for the associated processing. Furthermore, a distributed architecture enables performing the computations close to the antenna, possibly integrated on the same chip as the radio. 
In the case of component failure, the modularity allows a single node to be replaced instead of replacing a large centralized unit.

Additionally, for centralized implementations the required data rate to read all uplink data from the ADCs and to feed the downlink data to the DACs grows with the number of antennas, making it very high for systems with many antennas. Finally, a higher manufacturing yield can be expected since each chip is smaller for a distributed architecture. 

In the current work, a node and system architecture is proposed that is distributed, modular, and scalable. It supports conjugate beamforming (CB), zero forcing (ZF)  \cite{Yang2013}, and minimum mean-square error (MMSE) \cite{peel2005vector}, where in the two latter cases a matrix inversion is performed in a central unit. The computational difference between ZF and MMSE is that in MMSE, a regularization term is added before performing the matrix inverse. This is also done at the central unit. Therefore, we only discuss CB and ZF explicitly, as the node processing for MMSE is the same as for ZF.
The main contributions in this work are:
\begin{itemize}
	\item Analysis of distributed and modular processing in a massive MIMO-OFDM system
	\item Node and system architecture for a distributed, modular, and scalable MIMO-OFDM system
	\item Computation, memory, and communication analysis for the nodes and system
	\item Analysis of timing constraints and their effects on resource requirements
	\item Deterministic scheduling/control of the nodes and system
	\item Design space exploration showing that the proposed node architecture can be used in a rather large set of scenarios
\end{itemize}

A preliminary version of the current manuscript was presented in \cite{Bertilsson2016}. Compared to the distributed architecture in \cite{Puglielli2016}, we suggest using a tree interconnection of the nodes, although the proposed approach can also be used in other interconnection topologies. Especially, we perform an analysis of the computational and timing requirements and propose a detailed node architecture, along with scheduling of the computations and inter-node communication. Compared to the distributed architectures in \cite{kaipeng2016distributed, li2016decentralized}, we propose an optimized node architecture instead of using generic processors. Additionally, the timing constraints of the selected frame format is carefully analyzed.
\section{Proposed System Architecture}\label{sec:system_architecture}

\begin{figure}[tb]
		\centering
			\includegraphics[scale=0.5]{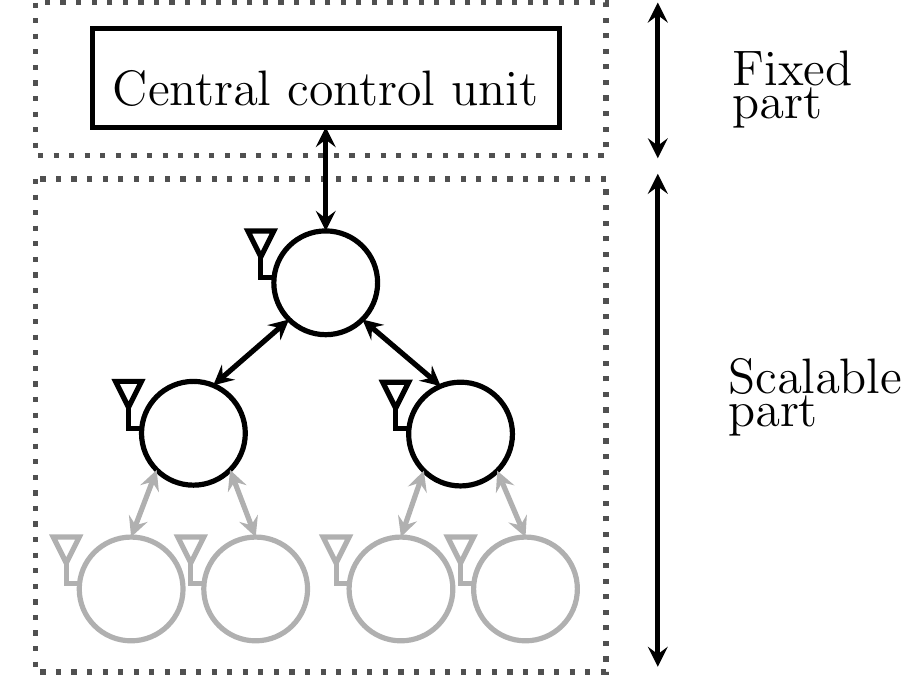}
		\caption[]{\label{fig:overview} Proposed system architecture consisting of a central control unit (CCU) and the scalable antenna nodes.}
	\end{figure}

	In this work, the proposed system architecture consists of one central control unit (CCU) and a scalable part, as illustrated in Fig.~\ref{fig:overview}. The CCU is responsible for performing operations such as error correction coding/decoding and operations associated with the other network layers, such as medium access control (MAC). The scalable part is responsible for the channel estimation, linear precoding, and linear decoding of symbols transmitted to and from the base station. Every node in the scalable part contains computational blocks for the associated antenna(s) and inter-node communication links. One or more nodes can be combined into a chip for different granularity. The main difference is the latency of the inter-node communication, which within a chip will be one or a few clock cycles, while inter-chip may be in the range of one hundred clock cycles assuming a serial link and a clock frequency of hundreds of MHz. Here it is assumed that the nodes are clocked synchronously. In downlink operation, the CCU feeds modulated symbols for each terminal to the nodes. In uplink operation, the nodes compute estimated symbols transmitted from the terminals and sends to the CCU.
	
\begin{figure}[tb]
		\centering
			\includegraphics[scale=0.3]{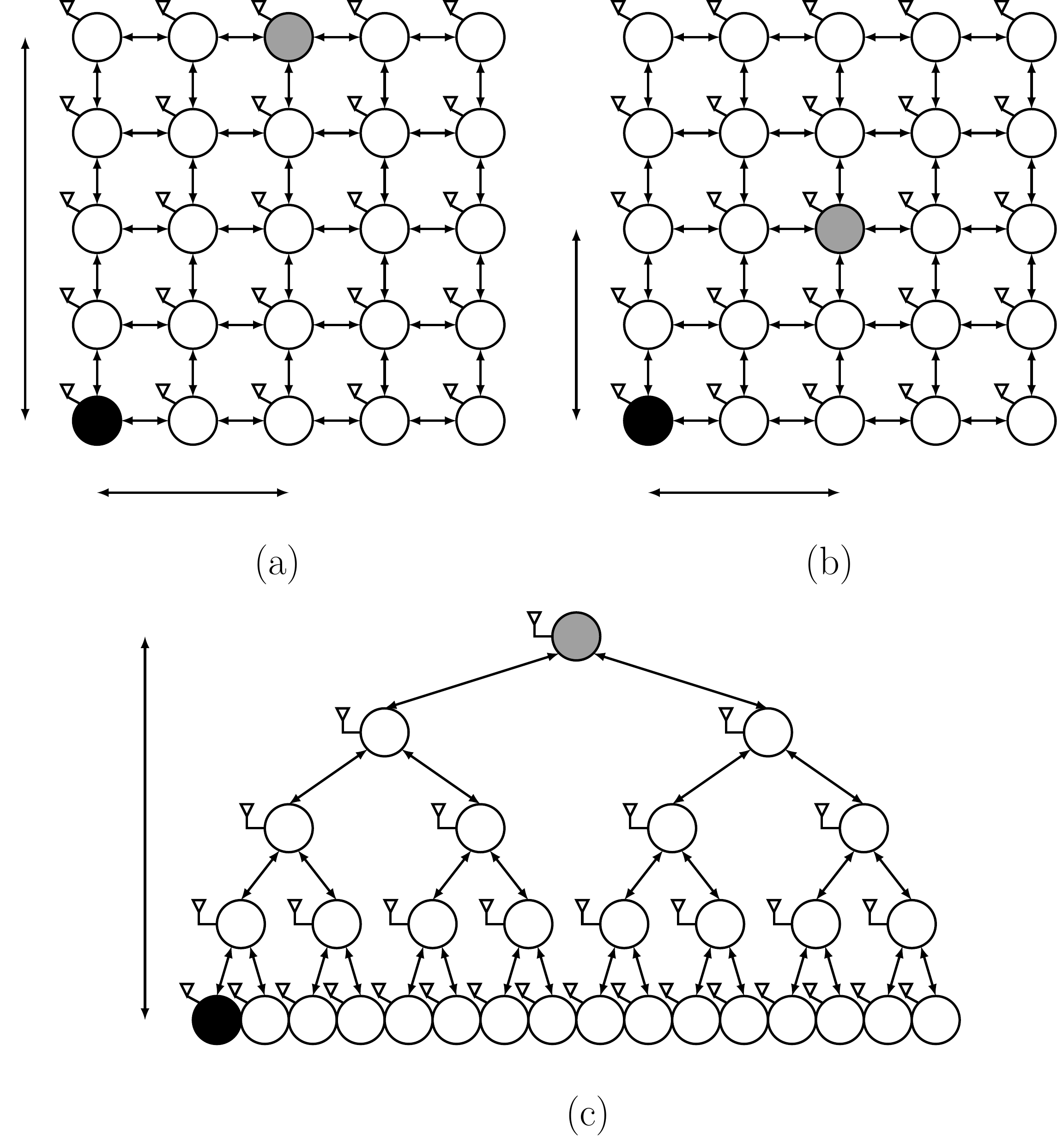}
		\caption[]{\label{fig:array} Three different node topologies. The external controller is connected to the grey nodes. The black node is (one of) the furthest. The number of routing hops is: (a) $N_{\text{hops}} \propto \frac{3}{2}\sqrt{M}$, (b) $N_{\text{hops}} \propto \sqrt{M}$, (c) $N_{\text{hops}} \propto \log_2M$.}
	\end{figure}

	In this work, we propose to connect the nodes in a $K$-ary tree, however the nodes can be connected in other topologies as well.	It is worth pointing out that, independently of the interconnection topology, during the accumulation of data from all nodes, each node will only transmit data to one other node on the way to the CCU to avoid duplicate transmissions and accumulations. This means that accumulating data will always be performed in a tree structure, independently of the interconnection topology. By modifying the interconnection topology, the number of hops when accumulating data is changed. Trees have some inherent advantages and disadvantages compared to array topologies.	One of the most profound advantages of the tree structure is that the number of routing hops, $N_{\text{hops}}$, grows logarithmically with the number of nodes in the tree, as opposed to proportionately to the square root for arrays. Figure~\ref{fig:array} shows two different arrays and a tree topology. For systems with a large number of antennas this is a major benefit when using ZF or MMSE processing, as the latency of propagating data through the tree affects the system design. For CB processing, low latency is not as important.

	Another design trade-off that needs to be considered is fault tolerance. In an ordinary tree structure, if a node fails during operation that entire branch will not be able to communicate with the rest of the tree. In an array topology, this could be mitigated by routing data past the failing node. This however increases the node complexity since a routing mechanism must be implemented. 

	Additionally, there is the aspect of physical antenna placement and cable routing. In systems where antennas are placed in an array, the array-based node topologies have the advantage of simpler cable routing. In systems where the antennas are scattered in some irregular pattern this advantage is lost.

	For the remainder of the article, for ease of exposition, complete binary trees are considered where each chip contains one node.

	\subsection{System Specification}\label{sec:system_specification}
\begin{figure}[tb]
			\centering
				\includegraphics[scale=0.65]{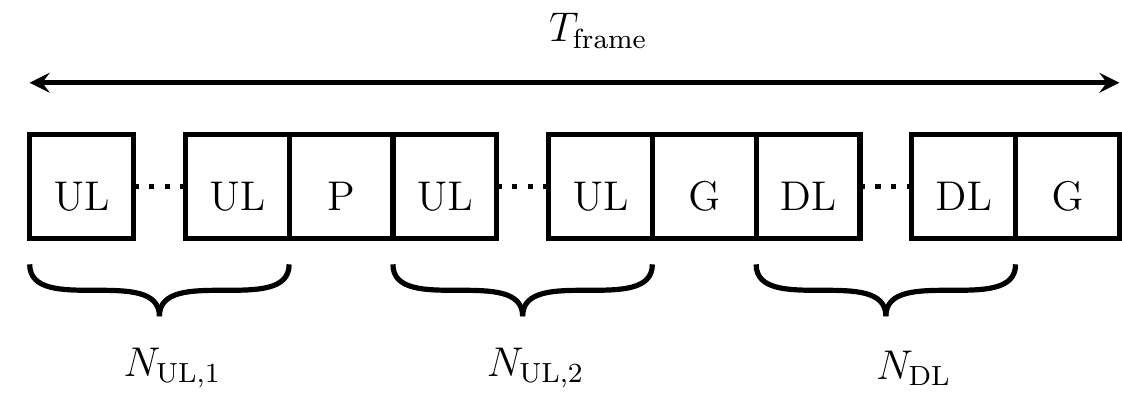}
			\caption[]{\label{fig:frame} Generalized frame structure.}
		\end{figure}

	The considered setup is a TDD based system that utilizes OFDM. A generalized frame structure can be seen in Fig.\,\ref{fig:frame}. The frame starts with $N_{\text{UL,1}}$ uplink OFDM symbols where the terminals transmits data to the base station. Then comes the uplink pilot symbol, where all terminals transmit a unique pilot sequence that is used to estimate the uplink radio channel. Another $N_{\text{UL,2}}$ uplink OFDM symbols are sent after the pilot. Then comes a guard interval to switch from uplink to downlink operation. The base station then transmits $N_{\text{DL}}$ OFDM symbols to the terminals. The frame duration is
	\begin{equation}
		T_{\text{frame}}=
		\left(N_{\text{UL,1}}+N_{\text{UL,2}}+N_{\text{DL}}+3\right)T_{\text{OFDM}},\label{eq:tframe}
	\end{equation}
	where $T_{\text{OFDM}}$ is the duration of one OFDM symbol.
		
	Generally, it is favorable to place the pilot close to the middle of the frame to reduce the time between sending the pilot and the data.

	Here, the synchronization between transmitters and receivers is not considered.
	The system parameters are shown in Table~\ref{tab:spec}.
	\begin{table}
		\centering
			\caption{System Parameters.}
			\label{tab:spec}
			\begin{tabular}{  l | l }
				\hline
				\textbf{Name} & \textbf{Description} \\ \hline
				$K$ & Number of terminals \\ 
				$N_{\text{FFT}}$ & OFDM DFT/IDFT lenght \\ 
				$N_{\text{SC}}$ & OFDM subcarriers utilized \\ 
				$N_{\text{UL,1}}$ & Uplink OFDM symbols before pilot \\
				$N_{\text{UL,2}}$ & Uplink OFDM symbols after pilot \\ 
				$N_{\text{DL}}$ & Downlink OFDM symbols \\ 
				$N_{\text{hops}}$ & Number of hops to the furthest node \\
				$f_{\text{sample}}$ & Sample rate \\
				$T_{\text{OFDM}}$ & Duration of one OFDM symbol \\ 
				$T_{\text{frame}}$ & Duration of one frame \\ 
				$T_{\text{inv}}$ & Time to compute matrix inverse \\
				$T_{\text{link}}$ & Latency of sending one value over the link \\ 		
				$W_{\text{comp}}$ & Word length of partial results\\ 
				$W_{\text{symbol}}$ & Word length of QAM modulated symbols \\ 
				$W_{\text{ADC}}$ & Word length of ADC\\
				$W_{\text{DAC}}$ & Word length of DAC \\ \hline
			\end{tabular}
	\end{table}

\section{Computational Tasks}\label{sec:mapping}
	To utilize the proposed architecture efficiently, the algorithms used must be expressed in a distributed manner. The processing can be divided into three phases: channel estimation, uplink data decoding and downlink data precoding.

	\subsection{Channel Estimation}
	Here, a channel estimation based on least squares is considered. Let
	\begin{equation}\label{eq:pilot}
			\mathbf{x}^k_{\text{pilot}}=
			\begin{bmatrix}
				0_{1 \times (k-1)} & p & 0_{1 \times (K-k)} \\
			\end{bmatrix}
	\end{equation}
	be the pilot vector transmitted by terminal $k$. The scalar $p$ is computed statically at design time. Each node receives the signal vector
	\begin{equation}
		\mathbf{y}_{\text{i,pilot}}=
		\mathbf{h}_i
		\mathbf{X}_p+
		\mathbf{N}_p
		\in\mathbb{C}^{1 \times K},
	\end{equation}
	where $\mathbf{X}_p\in\mathbb{C}^{K \times K}$ is the pilot matrix and $\mathbf{N}_p\in\mathbb{C}^{1 \times K}$ is a noise vector. The pilot matrix $\mathbf{X}_p$ is given by
	\begin{equation}
		\mathbf{X}_p=
		\begin{bmatrix}
			\left(\mathbf{x}^1_{\text{pilot}}\right)^{\intercal} & \left(\mathbf{x}^2_{\text{pilot}}\right)^{\intercal} & \cdots & \left(\mathbf{x}^k_{\text{pilot}}\right)^{\intercal} 
		\end{bmatrix}.
	\end{equation}
	When the pilot signals has been received at node $i$, it has all data necessary data to estimate the channels to the $K$ users, without any inter-node communication. This is done by multiplying the received pilot signals by the scalar $\frac{1}{p}$. The local channel estimate vector is
	\begin{equation}\label{eq:ce}
		\hat{\mathbf{h}_i} = \frac{\mathbf{y}_{\text{i,pilot}}}{p} \in\mathbb{C}^{1 \times K}.
	\end{equation}
	Assuming that the channels are frequency flat, the entire channel estimate matrix $\mathbf{H}\in\mathbb{C}^{M \times K}$ can be written as
	\begin{equation}
		\mathbf{H} =
		\begin{pmatrix}
			h_{1,1} & h_{1,2} & \cdots & h_{1,K} \\
			h_{2,1} & h_{2,2} & \cdots & h_{2,K} \\
			\vdots  & \vdots  & \ddots & \vdots  \\
			h_{M,1} & h_{M,2} & \cdots & h_{M,K} 
		\end{pmatrix},
	\end{equation}
	where $h_{\text{i,j}} \in \mathbb{C}$ is the channel coefficient between antenna $i$ and terminal $j$. After the locally performed channel estimation, node $i$ has computed and stored row $i$ of the channel matrix.
	
	


	\subsection{Linear Decoding and Precoding Matrices}\label{sec:dpm}
	
	In the uplink data transmission, the base station separates the received signal vector $\mathbf{y}\in\mathbb{C}^{M \times 1}$ into $K$ streams of symbols $\mathbf{\tilde{y}}\in\mathbb{C}^{K \times 1}$. This is done by multiplication with a linear detection matrix $\mathbf{A}\in\mathbb{C}^{K \times M}$. For the considered algorithms, the decoding matrix is
	\begin{equation}
		\mathbf{A} = \left\{
			\begin{array}{lr}
				\mathbf{H}^H\textrm{, for CB}\\
				\left(\mathbf{H}^H\mathbf{H}\right)^{-1}\mathbf{H}^H\text{, for ZF.}
			\end{array}
		\right.
	\end{equation}
	
	In the downlink data transmission, the symbol vector $\mathbf{q}\in\mathbb{C}^{K \times 1}$ is precoded and sent from $M$ antennas $\mathbf{x}\in\mathbb{C}^{M \times 1}$. This is done by multiplication with a linear precoding matrix $\mathbf{W}\in\mathbb{C}^{M \times K}$. For the considered algorithms, the precoding matrix is
	\begin{equation}
		\mathbf{W} = \left\{
			\begin{array}{lr}
				\mathbf{H}^\ast\text{, for CB}\\
				\mathbf{H}^\ast\left(\mathbf{H}^{\intercal}\mathbf{H}^\ast\right)^{-1}\text{, for ZF.}
			\end{array}
		\right.
	\end{equation}

	For CB, the linear detection matrix $\mathbf{A}$ and the linear precoding matrix $\mathbf{W}$ are obtained directly from the channel estimation. Each node then has access to one column of the decoding matrix.
	
	For the ZF algorithm, calculating $\mathbf{A}$ and $\mathbf{W}$ involves performing a pseudo inverse of the channel matrix $\mathbf{H}$. The ZF precoding matrix is
	\begin{equation}
		\mathbf{W} = 
		\mathbf{H}^\ast\left(\mathbf{H}^{\intercal}\mathbf{H}^\ast\right)^{-1}=
		\mathbf{H}^\ast\left(\left(\mathbf{H}^H\mathbf{H}\right)^{-1}\right)^\ast.
	\end{equation}
	Let 
	\begin{equation}
		\mathbf{D} = \left(\mathbf{H}^H\mathbf{H}\right)^{-1}.
	\end{equation}
	The matrices can for the ZF case then be rewritten as
	\begin{equation}
		\mathbf{W} =
		\mathbf{H}^\ast\mathbf{D}^\ast=
		\left(\mathbf{H}\mathbf{D}\right)^\ast
	\end{equation}
	and
	\begin{equation}
		\mathbf{A}=
		\mathbf{D}\mathbf{H}^H.
	\end{equation}

	Given the fact that $\mathbf{H}^H\mathbf{H}$ is Hermitian, we know that its inverse is also Hermitian. With the Hermitian property ($\mathbf{D}=\mathbf{D}^H$), the decoding matrix can be written as
	\begin{equation}
		\mathbf{A}=
		\mathbf{D}^H\mathbf{H}^H=
		\left(\mathbf{H}\mathbf{D}\right)^H=\mathbf{W}^{\intercal}.
	\end{equation}

	Since the decoding and precoding matrices are each others transpose, the local decoding column vector and the precoding row vector will be identical.

	To calculate $\mathbf{A}$ and $\mathbf{W}$, $\mathbf{D}$ must be known. The Gram matrix of the channel estimates, $\mathbf{H}^H\mathbf{H}$, can be calculated in a distributed manner across all nodes. The inversion is then performed in the CCU. Let $\mathbf{B}=\mathbf{H}^H\mathbf{H}$.
		\begin{IEEEeqnarray}{rCl}
			\IEEEeqnarraymulticol{3}{l}{\mathbf{H}^H\mathbf{H}= \sum_{i=1}^{M}\mathbf{h}_i^H\mathbf{h}_i} \nonumber \\
			\ \ & =&
			\begin{pmatrix}
				h_{1,1}^\ast & \cdots & h_{M,1}^\ast \\
				h_{1,2}^\ast & \cdots & h_{M,2}^\ast \\
				\vdots  & \ddots & \vdots  \\
				h_{1,K}^\ast & \cdots & h_{M,K}^\ast 
			\end{pmatrix}
			\begin{pmatrix}
				h_{1,1}^\ast & h_{1,2}^\ast & \cdots & h_{1,K}^\ast \\
				\vdots & \vdots & \ddots & \vdots  \\
				h_{M,1}^\ast & h_{M,2}^\ast & \cdots & h_{M,K}^\ast
			\end{pmatrix} \nonumber \\
		\end{IEEEeqnarray}
		The matrix $\mathbf{h}_i^H\mathbf{h}_i$ is the Gram matrix of the local channel estimate vector in node $i$, and can be computed locally without any inter-node communication since the required data is obtained from the channel estimation. It is a Hermitian matrix, thus only $\frac{K(K+1)}{2}$ entries must be computed. The computation performed in node $i$ is
		\begin{equation}\label{eq:bi}
			\mathbf{B}_i=\mathbf{h}_i^H\mathbf{h}_i+\mathbf{B}_{\text{left child}}+\mathbf{B}_{\text{right child}}.
		\end{equation}
		The local contributions are added together as the matrices are propagated upwards in the tree to form the Gram matrix of the channel estimates.
	
	This reduces the computational complexity of the CCU and reduces the amount of data to be sent in the tree. Instead of propagating a matrix with $M \times K$ values, due to the Hermitian property only $\frac{K(K+1)}{2}$ values needs to be propagated. However, the computational load in each node is increased, since $\mathbf{B}_i$ must be computed at the node.

	When the $\mathbf{D}$ matrix has been computed in the CCU, it is propagated downwards in the tree structure to all nodes. The nodes can then calculate their local detection and precoding vectors by multiplying the inverted matrix with their local channel estimate vector. The computation performed in node $i$ is
	\begin{equation}\label{eq:aw}
		\mathbf{A}_i=
		\mathbf{D}\mathbf{h}_i
		\in \mathbb{C}^{K \times 1},
	\end{equation}
	where $\mathbf{A}_i$ is the local decoding vector and $\mathbf{W}_i=\mathbf{A}_i^\intercal$ is the local precoding vector.
	
\begin{figure}
		\centering
			\includegraphics[scale=0.71]{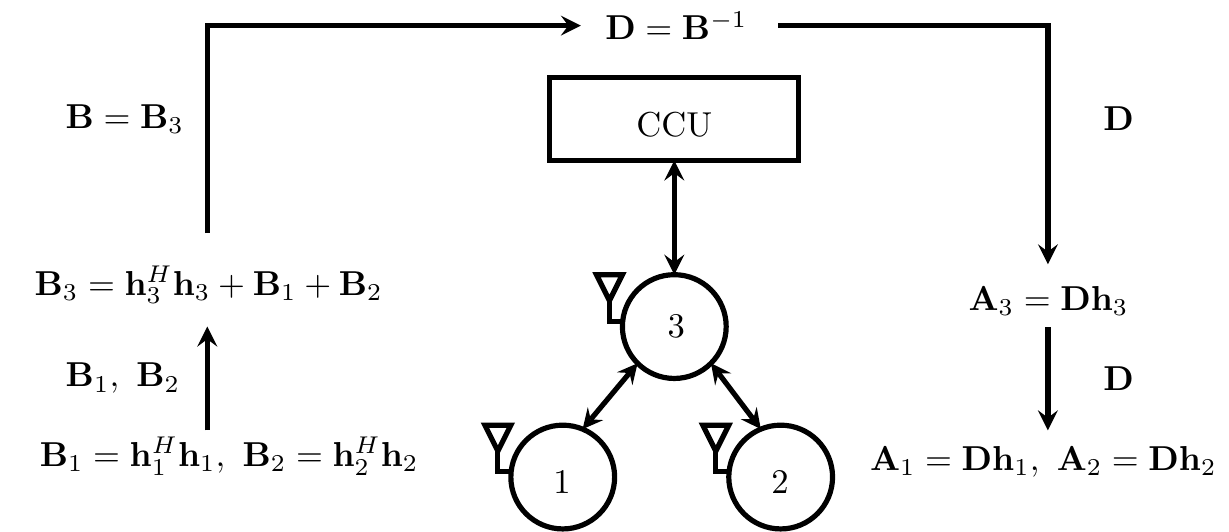}
		\caption[]{\label{fig:gram_local} Data transfers and partitioning of computations when determining the precoding/decoding vector for $M=3$.}		 
	\end{figure}

	The process of determining the local precoding/decoding vector, $\mathbf{A}_i$, is illustrated in Fig.~\ref{fig:gram_local}. The leaf nodes, $1$ and $2$, computes their local contribution to the Gram matrix, $B_1$ and $B_2$ respectively, and sends them to the parent node, $3$. Node $3$ computes its own local contribution, $B_3$, and sums it together with the contributions from the child nodes before sending it upwards to the CCU. The CCU performs the matrix inversion, and redistributes the results downwards in the tree. When each node receives the inverted matrix, it computes its local precoding/decoding vector.
	
	\subsection{Uplink Linear Decoding}\label{sec:ULD}
	The decoding process is performed by multiplying the received signal vector $\mathbf{y}\in\mathbb{C}^{M \times 1}$ with the decoding matrix, $\mathbf{A}$. During the decoding, each node has access to one column of the decoding matrix and one sample of the received signal vector. The symbol vector estimate $\tilde{\mathbf{y}}$ is
		\begin{IEEEeqnarray}{rCl}
		\tilde{\mathbf{y}} & = & \mathbf{A}\mathbf{y} = 
		\begin{pmatrix}
			a_{1,1} & a_{1,2} & \cdots & a_{1,M} \\
			a_{2,1} & a_{2,2} & \cdots & a_{2,M} \\
			\vdots  & \vdots  & \ddots & \vdots  \\
			a_{K,1} & a_{K,2} & \cdots & a_{K,M} 
		\end{pmatrix}
		\begin{pmatrix}
			y_{1}\\
			y_{2}\\
			\vdots\\
			y_{M}
		\end{pmatrix}\nonumber \\
		& = & 
		\begin{pmatrix}
			a_{1,1} y_{1} \\
			a_{2,1} y_{1} \\
			\vdots\\
			a_{K,1} y_{1} \\
		\end{pmatrix}
		+
		\begin{pmatrix}
			a_{1,2} y_{2} \\
			a_{2,2} y_{2} \\
			\vdots\\
			a_{K,2} y_{2} \\
		\end{pmatrix}
		+
		\hdots
		+
		\begin{pmatrix}
			a_{1,M} y_{M} \\
			a_{2,M} y_{M} \\
			\vdots\\
			a_{K,M} y_{M} \\
		\end{pmatrix}.
	\end{IEEEeqnarray}
	By multiplying the local sample with the local decoding column, a local contribution to the received symbol vector is computed. When the local contributions are calculated in each node, they are sent upwards in the tree structure. The contributions are added together as they propagate to the CCU. Since the local contribution can be calculated using only the local sample and one column of the decoding matrix, the entire decoding matrix does not need to be available in all nodes. The computation performed for each subcarrier in node $i$ is
	\begin{equation}\label{eq:yi}
		\tilde{\mathbf{y}}_i=
		\mathbf{A}_iy_i+\tilde{\mathbf{y}}_{\text{left child}}+\tilde{\mathbf{y}}_{\text{right child}},
	\end{equation}
	where $\mathbf{A}_i$ is the local decoding vector. This is computed similarly to computing $\mathbf{B}$ in Fig.~\ref{fig:gram_local}.

	
%

	\subsection{Downlink Linear Precoding}\label{sec:DLP}
	The precoding process is done by multiplying the symbol vector, $\mathbf{q}\in\mathbb{C}^{K \times 1}$, with the precoding matrix, $\mathbf{W}\in\mathbb{C}^{M \times K}$. During the precoding, each node has access to the symbol vector and one row of the precoding matrix.

\begin{IEEEeqnarray}{rCl}
		\mathbf{x}&= & \mathbf{W}\mathbf{q}=
		\begin{pmatrix}
			w_{1,1} & w_{1,2} & \cdots & w_{1,K} \\
			w_{2,1} & w_{2,2} & \cdots & w_{2,K} \\
			\vdots  & \vdots  & \ddots & \vdots  \\
			w_{M,1} & w_{M,2} & \cdots & w_{M,K} 
		\end{pmatrix}
		\begin{pmatrix}
			q_1 \\
			q_2 \\
			\vdots \\
			q_K \\
		\end{pmatrix} \nonumber \\
		& = &	
		\begin{pmatrix}
			\sum_{j=1}^{K} \mathbf{W}_{1,j} q_j \\
			\sum_{j=1}^{K} \mathbf{W}_{2,j} q_j \\
			\vdots \\
			\sum_{j=1}^{K} \mathbf{W}_{M,j} q_j \\
		\end{pmatrix}.
	\end{IEEEeqnarray}
	
	The value transmitted at node $i$ is the inner product between row $i$ of the precoding matrix and the symbol vector $\mathbf{q}$. Similarly to the decoding case, each node only requires one row of the precoding matrix to perform the precoding. Thus, the entire matrix does not need to be distributed to all nodes. The computation performed for each subcarrier in node $i$ is
	\begin{equation}\label{eq:xi}
		\mathbf{x}_i=
		\sum_{j=1}^{K} \mathbf{W}_{i,j} q_j,
	\end{equation}
	where $\mathbf{W}_i$ is the local precoding vector. The symbol vector, $\mathbf{q}$, is distributed to the nodes similarly to $\mathbf{D}$ and the computations of $\mathbf{x}_i$ is performed similarly to $\mathbf{A}_i$ in Fig.~\ref{fig:gram_local}.
	
	\subsection{OFDM Modulation and Demodulation}\label{sec:ofdm_modulation}
	In a massive MIMO OFDM system, the OFDM modulation and demodulation is performed for each antenna. Therefore, one FFT/IFFT must be performed in the node for each OFDM symbol (pilot, uplink, and downlink). The length of the FFT/IFFT is $N_{\text{FFT}}$, while the number of subcarriers utilized is $N_{\text{SC}}$.
	
	\subsection{Processing Element}\label{sec:processingelement}
\begin{figure}
			\centering
				\resizebox{\linewidth}{!}{
					\includegraphics{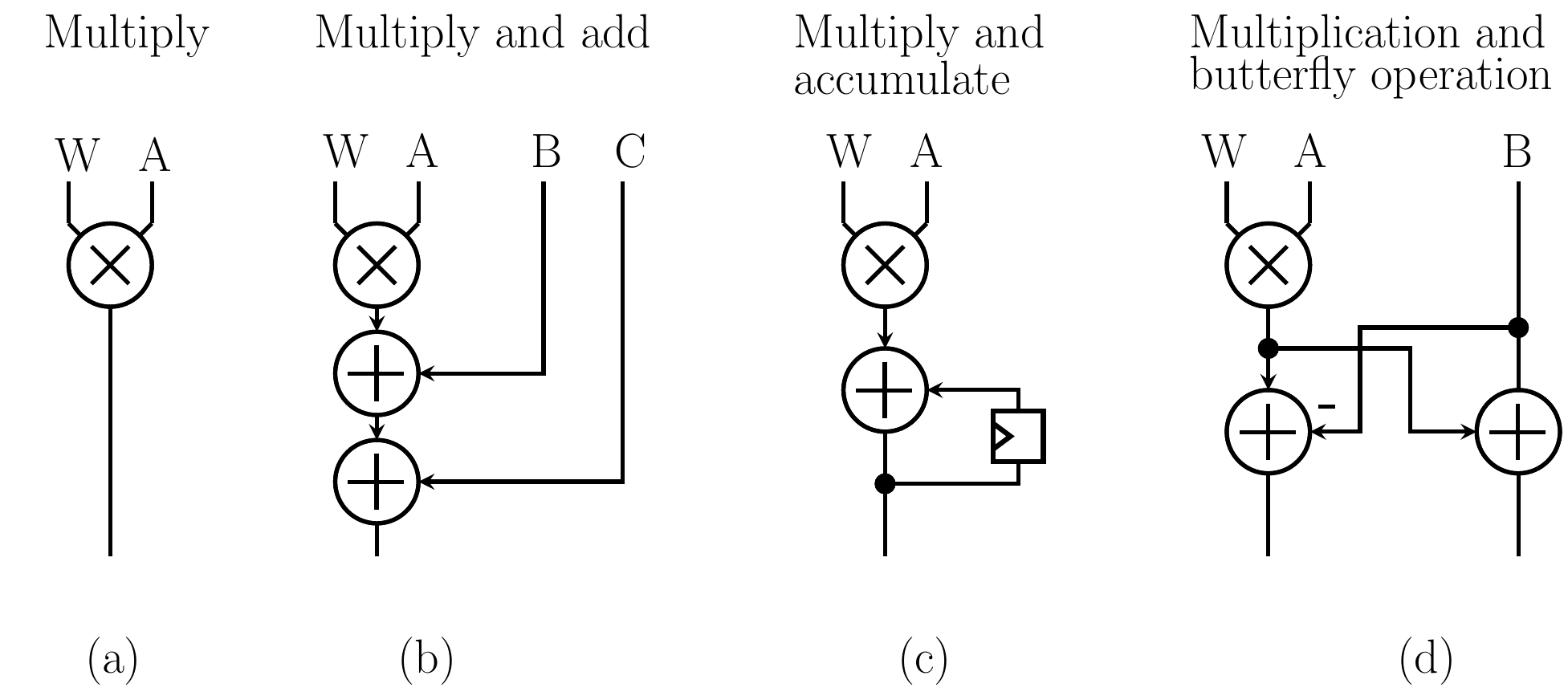}
				}
			\caption[]{\label{fig:arith_op} Arithmetic operations performed in the nodes.}
			 
		\end{figure}

	As is shown in Section~\ref{sec:dse}, having one processing element that performs all computations in the node is enough to support a large range of different combinations of the number of terminals and channel bandwidth. Therefore, it is beneficial to find a common structure of the involved computations discussed earlier. The channel estimation only requires multiplications with $1/p$ as shown in Fig.~\ref{fig:arith_op}(a). For uplink decoding, each node performs a multiplication and adds data from the other nodes further down the tree, for a binary tree as shown in Fig.~\ref{fig:arith_op}(b). For the downlink precoding, a sum-of-products is locally computed, which consists of multiplication and accumulation, as shown in Fig.~\ref{fig:arith_op}(c). 
	
	Finally, the FFT and IFFT consists of butterfly operations and twiddle factor multiplications. Considering the operations in Figs.~\ref{fig:arith_op}(a)--(c), it makes sense to use a radix-2 decimation in time (DIT) algorithm. This algorithm has the property that each butterfly operation has a twiddle factor multiplication in front of one of the inputs\cite{Wanhammar1999}, as shown in Fig.~\ref{fig:arith_op}(d). Although there exist many other radix-2 algorithm, the radix-2 DIT algorithm  is the only one with this property for each and every butterfly. As a note, it is often believed that DIT corresponds to bit-reversed input order and normal output order. However, this is not the case as the butterfly computation order, and, hence, the data dependency, is independent of the algorithm selection. A conflict-free memory access scheme with low hardware overhead can be found in e.g. \cite{Ma2000}.
	
\begin{figure}
			\centering
				\includegraphics[scale=0.5]{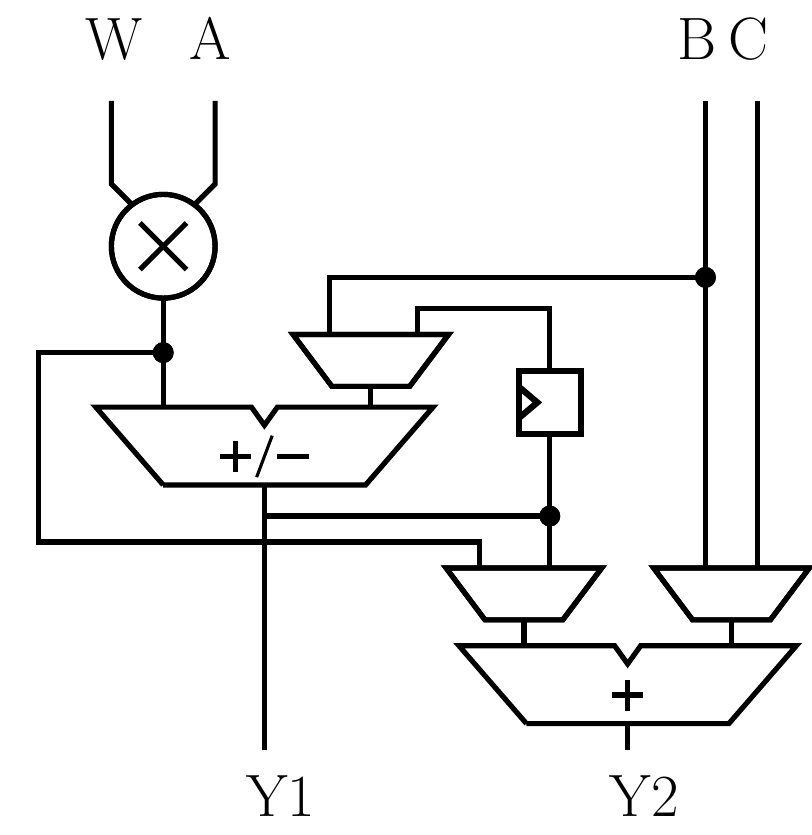}
			\caption[]{\label{fig:selected_pe} Proposed processing element capable of performing all the operations in Fig.~\ref{fig:arith_op}.}
		\end{figure}

		These operations can be efficiently mapped to a processing element as shown in Fig.~\ref{fig:selected_pe}. The number of operations for each task and the type of operations is summarized in Table~\ref{tab:num_op}.
	\begin{table}[tb]
		\centering
		\caption{Computational Tasks and the Number of Operations.}
		\label{tab:num_op}
			\begin{tabular}{  p{0.79cm} | p{2.64cm} | c | c  }
				\hline
				\textbf{Name} & \textbf{Description} & \textbf{\#PE Operations} & \textbf{Operation} \\ \hline
				CE & Channel estimation, (\ref{eq:ce}) & $K$ & Fig.~\ref{fig:arith_op}(a)\\ \hline
				$\mathbf{B}_i$ & Local contribution for Gram matrix, (\ref{eq:bi}) & $\frac{K(K+1)}{2}$ & Fig.~\ref{fig:arith_op}(b) \\ \hline
				$\tilde{\mathbf{y}}_{\text{i}}$ & Local $\tilde{\mathbf{y}}$ contribution and add contributions from child nodes, for all subcarriers, (\ref{eq:yi}) & $N_{\text{SC}}K$ & Fig.~\ref{fig:arith_op}(b) \\ \hline
				$x_i$ & Precoded symbol $x_i$ for all subcarriers, (\ref{eq:xi}) & $N_{\text{SC}}K$ & Fig.~\ref{fig:arith_op}(c)\\ \hline
				$\mathbf{W}_i$/$\mathbf{A}_i$ & Local precoding and decoding vector, (\ref{eq:aw}) & $K^2$ & Fig.~\ref{fig:arith_op}(c) \\ \hline
				FFT/ IFFT & FFT/IFFT for one OFDM symbol & $\frac{N_{\text{FFT}}}{2}\log_2\left(N_{\text{FFT}}\right)$ & Fig.~\ref{fig:arith_op}(d)\\ \hline
			\end{tabular}	
	\end{table}
	
	In cases where multiple processing elements are used, the processing element selection may be reconsidered. In this case, it might be beneficial to map different computational tasks to different processing elements, enabling specialized structures for the given task. Similarly, if the computational requirements per antenna are low, it may be beneficial to interleave the computations for more than one antenna on a single processing element.
	
	\subsection{Computational Partitioning}
	So far, all computations that can be performed is a distributed manner are assumed to be done so. However, this does not need to be the case. Consider ZF processing, where the uplink decoding is performed as
	\begin{equation}
		\tilde{\mathbf{y}}=
		\mathbf{A}\mathbf{y}=
		\left(\mathbf{H}^H\mathbf{H}\right)^{-1}\mathbf{H}^H\mathbf{y}.
	\end{equation}
	So far the decoding matrix, $\mathbf{A}=\left(\mathbf{H}^H\mathbf{H}\right)^{-1}\mathbf{H}^H$ is computed once every frame. This requires that the inverted matrix is redistributed to the nodes before the decoding process can start. Another possibility is to compute $\mathbf{H}^H\mathbf{y}$ in each node, just like the conjugate beamforming case, and multiply with the inverted matrix once the results reaches the CCU. The distributed parts of the decoding could then be started independently of the matrix inversion.
	
	Similarly for the downlink precoding, the ZF processing is performed according to
	\begin{equation}
		\mathbf{x}=
		\mathbf{W}\mathbf{q}=
		\mathbf{H}^\ast\left(\left(\mathbf{H}^H\mathbf{H}\right)^{-1}\right)^\ast\mathbf{q},
	\end{equation}
	where the precoding matrix $\mathbf{W}=\mathbf{H}^\ast\left(\left(\mathbf{H}^H\mathbf{H}\right)^{-1}\right)^\ast$ is computed once every frame. This requires that the inverted matrix is available in the node before the precoding can start. By multiplying the complex conjugate of the inverted matrix with the symbol vectors, $\left(\left(\mathbf{H}^H\mathbf{H}\right)^{-1}\right)^\ast\mathbf{q}$, in the CCU before they are sent to the nodes, the inverted matrix itself is not needed at each node for the precoding step. 
	
	By partitioning the computations in this way, the inverted matrix does not need to be redistributed to the nodes. However, the computational load in the CCU is significantly increased. The computational load of each node is only slightly reduced, since precoding and decoding are still performed distributedly. The only difference is that the $\mathbf{A}/\mathbf{W}$ computation does not need to be performed.

\section{Complexity Analysis}\label{sec:resource_allocation}
	In this section, the computational, memory, and communication complexity is analyzed.
	

%

	\begin{figure*}[tb]
		\centering
			\resizebox{\linewidth}{!}{
				\includegraphics{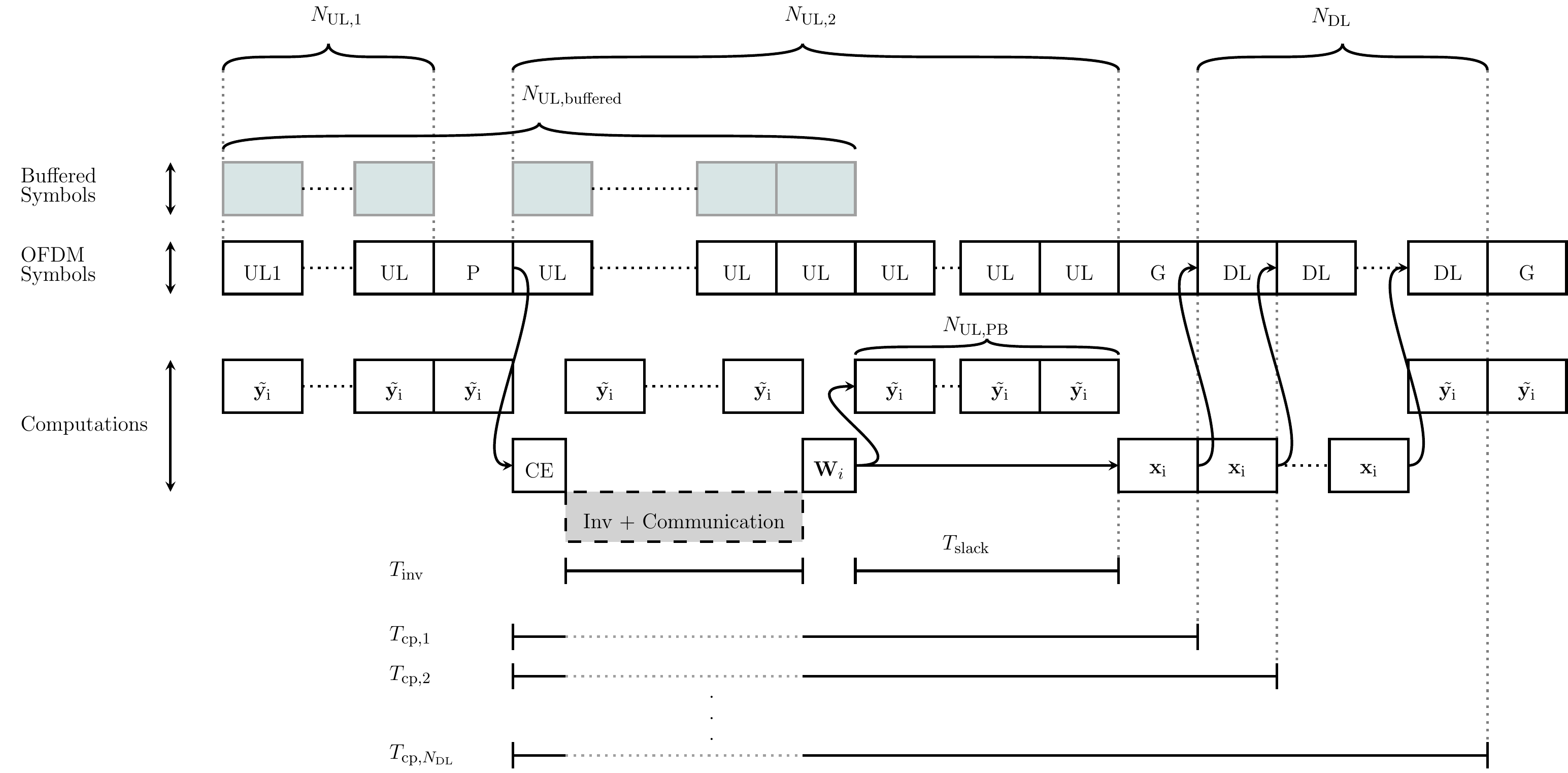}
			}
		\caption[]{\label{fig:buffer_size} Critical paths and their respective deadlines, for the asymptotic case $\left(N_{\text{OPS}}=N_{\text{OPS,asymptotic}}\right)$. Here, CE includes both channel estimation and computing $\mathbf{B}_i$. Dotted segments of the critical path times indicate that no computations on the critical path can be performed during the period. Gray boxes illustrate uplink OFDM symbols that must be stored before processing.}
	\end{figure*}
	
	\subsection{Computational Complexity}
	The computational complexity of each task is shown in Table~\ref{tab:num_op}. With the selected frame format, there are two major limitations on the computational resources. First, since the frame is repeated cyclically, all computations for one frame must be performed in the duration of one frame, $T_{\text{frame}}$. This yields the average number of operations per sample received, $N_{\text{op,avg}}$. The number of operations that needs to be performed to obtain the precoding/decoding vector is
	\begin{equation}
		N_{\text{op,weights}}=
		\frac{N_{\text{FFT}}}{2}\log_2\left(N_{\text{FFT}}\right)+
		K+
		\frac{K\left(K+1\right)}{2}+
		K^2,
	\end{equation}
	where the first term corresponds to demodulating the OFDM symbol (FFT), the second term for estimating the $K$ channels (CE) and the third term for computing the local contribution to the channel Gram matrix $\left(\mathbf{B}_i\right)$. The fourth term is from multiplying the inverted matrix with the local channel estimates to create the local decoding and precoding vector. The number of operations performed for each uplink OFDM symbol is
	\begin{equation}
		N_{\text{op,UL}}=
		\frac{N_{\text{FFT}}}{2}\log_2\left(N_{\text{FFT}}\right)+
		KN_{\text{SC}},
	\end{equation}
	where the first term corresponds to demodulating the OFDM symbol, and the second term for computing the local contribution to the received symbol vector. The number of operations required for each downlink OFDM symbol is
	\begin{equation}
		N_{\text{op,DL}}=
		KN_{\text{SC}}+
		\frac{N_{\text{FFT}}}{2}\log_2\left(N_{\text{FFT}}\right),
	\end{equation}
	where the first term corresponds to performing the precoding for each subcarrier utilized, and the second term for performing the OFDM modulation. The number of operations performed for the uplink and downlink OFDM symbols are the same:
	\begin{equation}
		N_{\text{op,OFDM}}=N_{\text{op,UL}}=N_{\text{op,DL}}.
	\end{equation}
	
	The total number of operations per sample on average over an entire frame is then
	\begin{equation}
		N_{\text{OPS,avg}}=
		\frac{
			N_{\text{op,weights}}+\left(N_{\text{UL}}+N_{\text{DL}}\right)N_{\text{op,OFDM}}
		}
		{
			T_{\text{frame}}f_{\text{sample}}
		},\label{eq:opsavg}
	\end{equation}
	where $N_{\text{UL}}=N_{\text{UL,1}} + N_{\text{UL,2}}$ is the total number of uplink OFDM symbols, and $N_{\text{DL}}$ is the number of downlink OFDM symbols. Without considering data dependencies or critical paths, this is the theoretical lower bound on the number of operations per sample that the node must be able to perform.

	The other limitation is that the downlink symbols must be processed before their respective deadlines. In practice there will be $N_{\text{DL}}$ critical paths in the schedule for one frame. Figure~\ref{fig:buffer_size} shows the computational tasks performed in each node, the critical paths in the frame, the important times and the possibility to buffer or process the uplink OFDM symbols. The critical paths in the computations are from receiving the pilot symbol, estimating the channels, computing the local contribution to the Gram matrix, performing the centralized matrix inversion, computing the local precoding/decoding vector, and finally performing the precoding for each downlink OFDM symbol. The number of operations on the critical path for downlink symbol $i$ is
	\begin{equation}
		N_{\text{op,CP,i}}=
		N_{\text{op,weights}}+iN_{\text{op,OFDM}},\  		i \in \{1,2,\hdots,N_{\text{DL}}\}.
		\label{eq:opcpi}
	\end{equation}
	The time available to perform the operations on the critical path for downlink symbol $i$ is
	\begin{equation}
		T_{\text{CP,i}}=
		T_{\text{OFDM}}\left(N_{\text{UL,2}}+i\right). \label{eq:tcpi}
	\end{equation}
	Between receiving the pilot symbol and transmitting downlink symbol $i$, there are $\left(N_{\text{UL,2}}+i\right)$ OFDM symbols, including the guard interval. However, during the time the local Gram matrices are propagated to the CCU, inverse computed and the result redistributed to the nodes, which in total takes
	$T_{\text{inv}}+2N_{\text{hops}}T_{\text{link}}$,
	no computations on the critical paths can be performed. Hence, the worst case average number of computations per sample on the critical paths is
	\begin{equation}
		N_{\text{OPS,critical}}=
		\underset{i}{\max}\left(
		\frac{
			N_{\text{op,CP,i}}
		}
		{
			\left(T_{\text{CP,i}}-T_{\text{inv}}-2N_{\text{hops}}T_{\text{link}}\right)f_{\text{sample}}
		}
		\right). \label{eq:opscritical}
	\end{equation}
	This leads to that the computational requirements are determined by
	\begin{equation}\label{eq:nops}
		N_{\text{OPS}}=\max\left(N_{\text{OPS,avg}}, N_{\text{OPS,critical}}\right).
	\end{equation}	
	This means that the time to perform matrix inversion and inter-node communication latency may affect the computational requirements.

	If the system specifications are kept, but the number of uplink and downlink OFDM symbols are increased, the average number of operations per sample over an entire frame increases as well. This is due to the two guard intervals becoming less significant with an increasing number of OFDM symbols. When the number of uplink and downlink OFDM symbols is large, the number of operations per sample is
	\begin{IEEEeqnarray}{rCl}
		N_{\text{OPS,asymptotic}} & = &
		\lim_{N_{\text{UL}},N_{\text{DL}} \to \infty}
		\max\left(N_{\text{OPS,avg}},N_{\text{OPS,critical}}\right) \nonumber
		\\
		&=&\frac
		{
			N_{\text{op,OFDM}}
		}
		{
			T_{\text{OFDM}}f_{\text{sample}} \label{eq:opsasymptotic}
		},
	\end{IEEEeqnarray}
	meaning that one OFDM symbol must be processed in the duration of one OFDM symbol.

\begin{figure}[tb]
		\centering
		\includegraphics[scale=0.6]{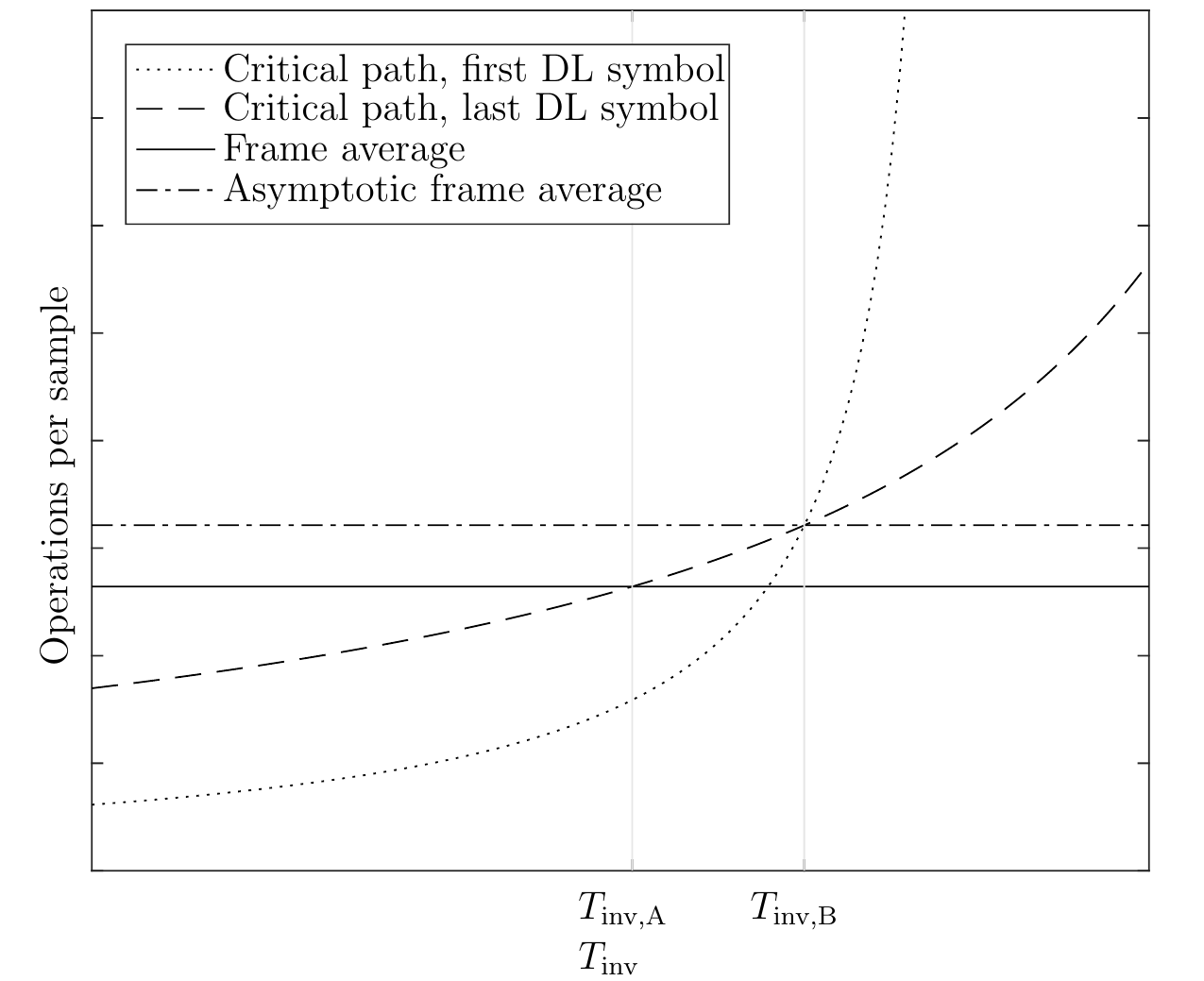}
		\caption[]{\label{fig:inv_time} Number of required PE operations per sample depending on the time to perform the matrix inversion.}
	\end{figure}

	As seen from (\ref{eq:opscritical}) and (\ref{eq:nops}), the matrix inversion time, $T_{\text{inv}}$, and the total inter-node communication latency, $2N_{\text{hops}}T_{\text{link}}$, may affect the computational requirements. For fixed inter-node communication latency\footnote{Note that the same behavior occurs when varying the inter-node communication latency, with fixed $T_{\text{inv}}$, or the sum of both.} this behavior is displayed in Fig.~\ref{fig:inv_time}.
		There are two inversion times marked in Fig.~\ref{fig:inv_time}. The first time, $T_{\text{inv,A}}$ is the time when the critical path requires equally many operations per sample as the frame average $\left(N_{\text{OPS,critical}} = N_{\text{OPS,avg}}\right)$. The second time, $T_{\text{inv,B}}$, is when the number of operations on the critical path grows larger than the number of operations per sample in the asymptotic case.
	
\begin{figure}[tb]
		\centering
			\includegraphics[scale=0.6]{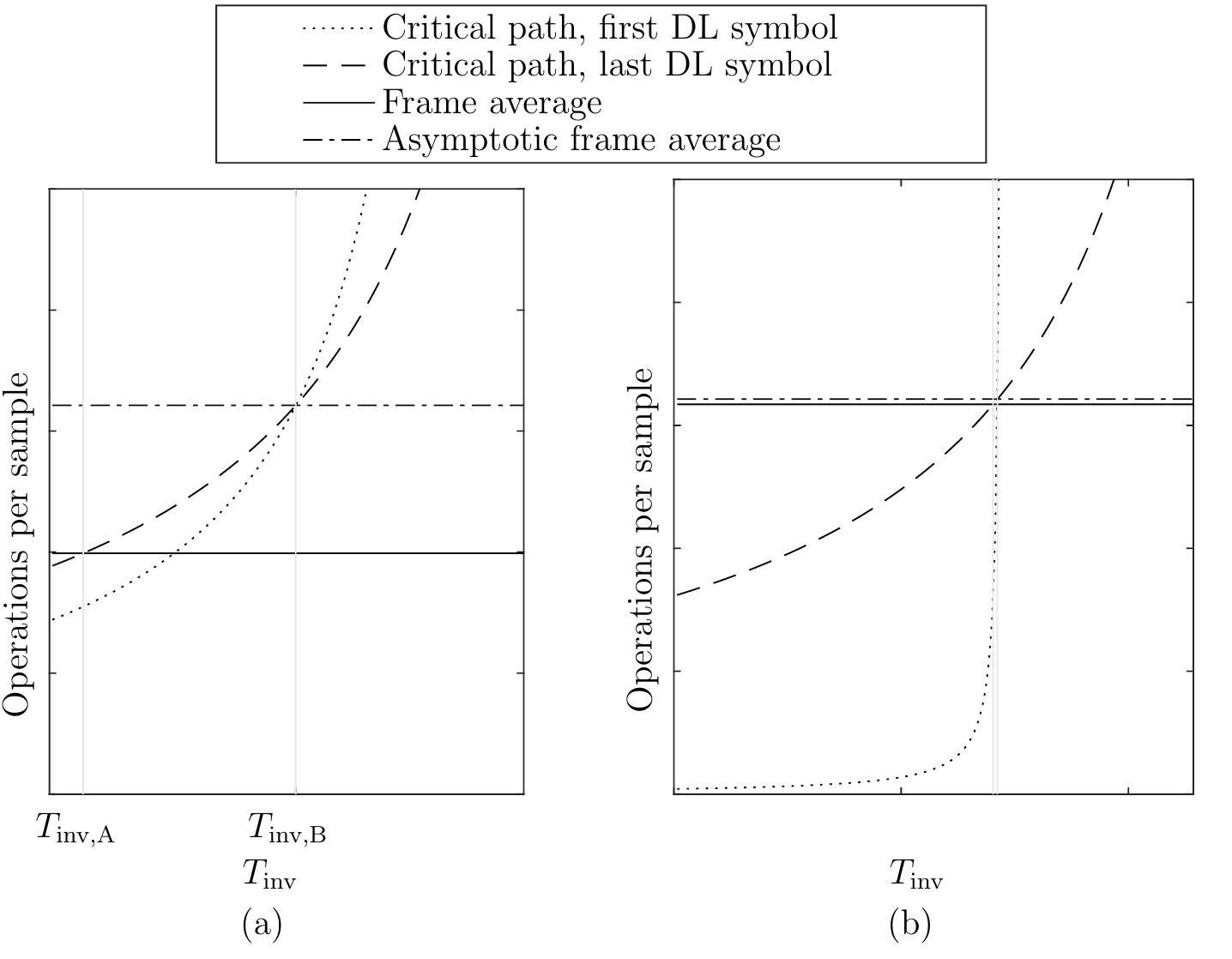}
		\caption[]{\label{fig:inv_time_cases} Number of operations per sample versus $T_{\text{inv}}$ for (a) small and (b) large number of OFDM symbols.}
	\end{figure}

	Figure \ref{fig:inv_time_cases} shows how the number of operations per sample for varying $T_{\text{inv}}$ changes when the number of OFDM symbols in a frame is changed. In Fig.~\ref{fig:inv_time_cases}(a) the number of OFDM symbols is small. In this case there is a significant gap between $N_{\text{OPS,avg}}$ and $N_{\text{OPS,asymptotic}}$ and between $T_{\text{inv,A}}$ and $T_{\text{inv,B}}$. When the number of OFDM symbols increases these gaps decreases, as shown in Fig.~\ref{fig:inv_time_cases}(b). Additionally, it can be seen in Fig.~\ref{fig:inv_time_cases}(b) that when the number of OFDM symbols is large, the time $T_{\text{inv,B}}$ acts as a deadline for the matrix inversion. If the inverse is received later than $T_{\text{inv,B}}$, the required number of operations per sample grows rapidly.
	
	
	It can be seen in Fig.~\ref{fig:inv_time_cases} that the critical path for the last downlink symbol is the first to cross the frame average line. Combining (\ref{eq:tframe}), (\ref{eq:opsavg}), (\ref{eq:tcpi}), and (\ref{eq:opscritical})  leads to 
	\begin{IEEEeqnarray}{rCl}
		\IEEEeqnarraymulticol{3}{l}{T_{\text{inv,A}} = 
			T_{\text{CP,}N_{\text{DL}}}-
			\frac{
				N_{\text{op,CP,}N_{\text{DL}}}
			}
			{
				N_{\text{op,total}}
			}
			T_{\text{frame}}-2N_{\text{hops}}T_{\text{link}}} \nonumber
		\\
		\ \ & = & T_\text{OFDM}\left(N_\text{UL,2}+N_{\text{DL}}\right. \nonumber \\
		&&\left.-\left(N_\text{UL}+N_{\text{DL}}+3\right)\frac{	N_{\text{op,weights}}+N_{\text{DL}}N_{\text{op,OFDM}}}{	N_{\text{op,weights}}+\left(N_{\text{UL}}+N_{\text{DL}}\right)N_{\text{op,OFDM}}}\right) \nonumber \\
		&& -2N_{\text{hops}}T_{\text{link}}.
	\end{IEEEeqnarray}
	
	The point $T_{\text{inv,B}}$ is given by the equation
	\begin{equation}
		N_{\text{OPS,asymptotic}}=
		\frac{
			N_{\text{op,CP,i}}
		}
		{
			\left(T_{\text{CP,i}}-T_{\text{inv,B}}-2N_{\text{hops}}T_{\text{link}}\right)f_{\text{sample}}
		},
	\end{equation}
	for downlink symbol $i$. Using (\ref{eq:opcpi}), (\ref{eq:tcpi}), and (\ref{eq:opsasymptotic}), $T_{\text{inv,B}}$ can be expressed as
	\begin{IEEEeqnarray}{rCl}
			T_{\text{inv,B}} 
			&= & T_{\text{OFDM}}
			\left(
				N_{\text{UL,2}}-
				\frac{
					N_{\text{op,weights}}
				}
				{
					N_{\text{op,OFDM}}
				}
			\right)-2N_{\text{hops}}T_{\text{link}}.
	\end{IEEEeqnarray}
	It can be seen that $T_{\text{inv,B}}$ is identical for all downlink symbols. The number of operations per sample for each of the critical paths are the same in the point $T_{\text{inv,B}}$. If $T_{\text{inv}} < T_{\text{inv,B}}$, the critical path to the last downlink symbol will always require the highest number of operations per sample of all critical paths. Similarly, if $T_{\text{inv}} > T_{\text{inv,B}}$ the critical path to the first downlink symbol requires the higher number of operations per sample.

	To keep up with the computational requirements, the number of operations per sample, $N_{\text{OPS}}$, the number of processing elements, $N_{\text{PE}}$, and the clock frequency, $f_{\text{clk}}$, must satisfy
	\begin{equation}\label{eq:pesfclk}
		N_{\text{PE}}\frac{f_{\text{clk}}}{f_{\text{sample}}}\geq
		N_{\text{OPS}} =
		\max\left(N_{\text{OPS,avg}},N_{\text{OPS,critical}}\right).
	\end{equation}
	Selecting $f_{\text{clk}}$ as an integer multiple of $f_{\text{sample}}$, the number of operations per sample that can be performed with $N_{\text{PE}}$ processing elements is
	\begin{equation}\label{eq:nopshat}
		\hat{N}_{\text{OPS}}=
		N_{\text{PE}}\frac{f_{\text{clk}}}{f_{\text{sample}}}.
	\end{equation}
	In most cases $N_{\text{OPS}}$ will not be an integer. However, $\hat{N}_{\text{OPS}}$ will, and, hence, there is a slack time that can be used to increase the number of terminals, $K$, the number of antennas, $M$, and/or the matrix inversion time, $T_{\text{inv}}$. If $T_{\text{inv}} < T_{\text{inv,A}}$, the slack time can be used to process some uplink symbols, say $N_{\text{UL,PB}}$, before the downlink symbols, as discussed below. Alternatively, the pilot symbol can be moved closer to the downlink symbols, i.e., decrease $N_{\text{UL,2}}$, as discussed in Section~\ref{sec:system_specification}.

	While this section focuses on ZF, the same analysis can be made for CB processing. This yields similar results, but with one significant difference. The precoding and decoding vector is obtained from the channel estimation, which means that the computational tasks $\mathbf{B}_i$, the central matrix inversion and $\mathbf{W}_i/\mathbf{A}_i$ is not performed. This results in
	\begin{equation}
		N_{\text{op,CP,i}}=
		\frac{N_{\text{FFT}}}{2}\log_2\left(N_{\text{FFT}}\right)+K+iN_{\text{op,OFDM}},\  		
		\label{eq:opcpicb}
	\end{equation}
	\begin{equation}
	N_{\text{OPS,critical}}=
	\underset{i}{\max}\left(
	\frac{
		N_{\text{op,CP,i}}
	}
	{
		T_{\text{CP,i}}f_{\text{sample}}
	}
	\right), \label{eq:opscriticalcb}
	\end{equation}
	and
	\begin{equation}
		N_{\text{OPS,avg}}=
		\frac{
			\frac{N_{\text{FFT}}}{2}\log_2\left(N_{\text{FFT}}\right)+K+\left(N_{\text{UL}}+N_{\text{DL}}\right)N_{\text{op,OFDM}}
		}
		{
			T_{\text{frame}}f_{\text{sample}}
		}\label{eq:opsavgcb}
	\end{equation}
	for CB. Hence, the number of operations to perform locally does not decrease significantly, but the latency issues of performing centralized computations vanishes.

	\subsection{Memory Complexity}
	Dimensioning the memories in the node will in part depend on the frame structure that is chosen, and in part on the scheduling of computations and inter-node communication. In Fig.~\ref{fig:buffer_size} the gray boxes illustrate uplink OFDM symbols that must be stored locally in the node before they are processed. The number of symbols that must be stored is
	\begin{equation}\label{eq:Nulbuffered}
		N_{\text{UL,buffered}}=
		N_{\text{UL}}-N_{\text{UL,PB}}
	\end{equation}
	and the number of bits required to store these symbols is
	\begin{equation}\label{eq:Nbitsbuffered}
		N_{\text{bits,buffered}}=
		N_{\text{UL,buffered}}N_{\text{FFT}}W_{\text{ADC}}.
	\end{equation}

\begin{figure}[tb]
		\centering
			\resizebox{\linewidth}{!}{
				\includegraphics{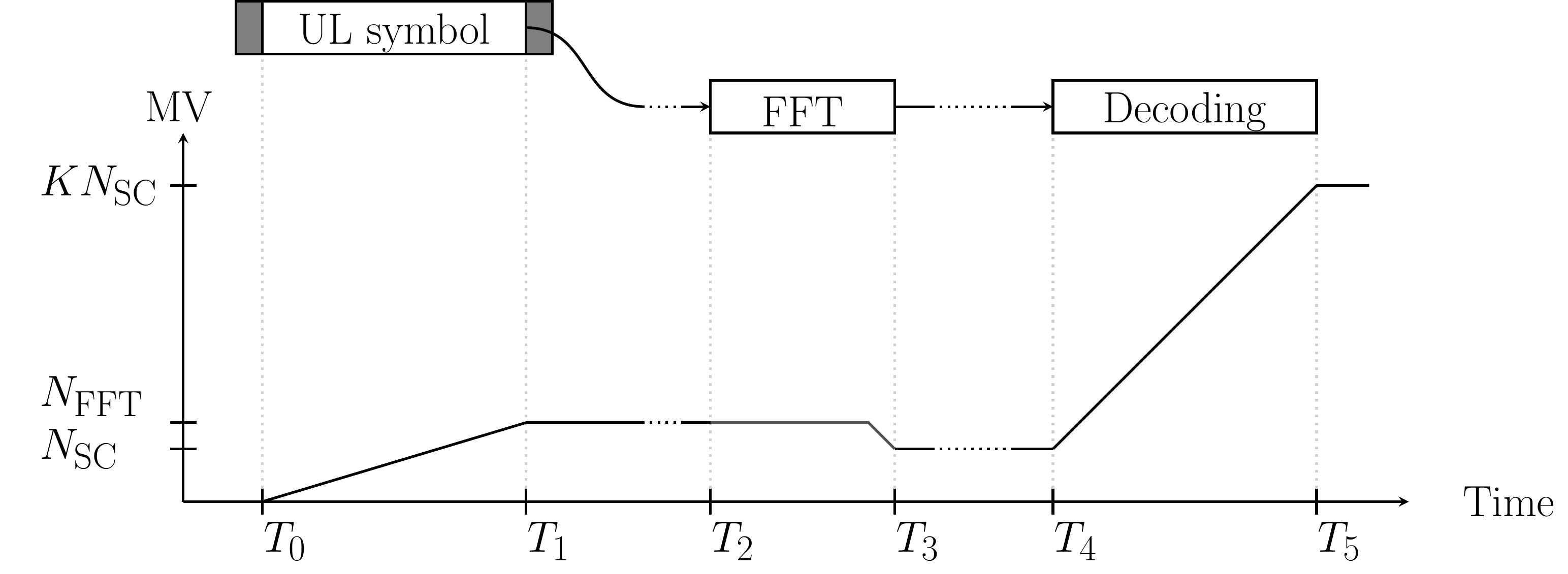}
			}
		\caption[]{\label{fig:ul_variables} Number of existing variables over the lifetime of an uplink OFDM symbol in the node.}
		
	\end{figure}

	For an uplink OFDM symbol, the number of variables during its lifetime in the node is seen in Fig.~\ref{fig:ul_variables}.
	Between times $T_0$ and $T_1$ the OFDM symbol is sampled from the antenna and stored in memory in the node. During this period, the number of variables grows until $N_{\text{FFT}}$. The duration between $T_0$ and $T_1$ is slightly shorter than one OFDM symbol, due to the cyclic prefix not being stored. At time $T_2$ the OFDM demodulation starts and is finished at time $T_3$. The FFT computation can be made in-place, meaning that no additional memory is strictly required. However, towards the end of the FFT computation, some variables can be discarded due to only $N_{\text{SC}}$ subcarriers being utilized. When the decoding starts at time $T_4$, there will be a data expansion by a factor $K$, since each subcarrier is multiplied with the decoding vector. When the decoding is finished, there are $KN_{\text{SC}}$ variables. These are the number of variables that will exist during the lifetime of one uplink symbol, but not all of them must be stored.

%
	
\subsection{Communication Complexity}
One of the advantages of distributing the computations among multiple nodes is that the number of values that needs to be sent to the centralized structure in the system grows proportionately to the number of terminals, $K$, rather than the number of antennas in the system, $M$. In massive MIMO systems where $M \gg K$ this is clearly advantageous.

	The number of bits that needs to be sent upwards in the tree structure during one frame is
	\begin{equation}
		N_{\text{bits,up}}=
		\left(\frac{K(K+1)}{2}+\left(N_{\text{UL,1}}+N_{\text{UL,2}}\right)N_{\text{SC}}\right)W_{\text{comp}},
	\end{equation}
	which corresponds to the local contributions to the Gram matrix, $\mathbf{B}_i$, and the symbol vector estimates, $\tilde{\mathbf{y}}_i$. These values are all used for computations and thus, the longer word length $W_{\text{comp}}$ is required. The required upwards link datarate is
	\begin{equation}
		R_{\text{up}}=
		\frac{
			N_{\text{bits,up}}
		}
		{
			T_{\text{frame}}
		}.
	\end{equation}

	The downwards propagation differs in that the word length of the modulated symbols is much shorter. Downwards, only the raw symbols are propagated to all nodes, using the shorter word length $W_{\text{symbol}}$. However, the inverted matrix still needs to be represented with $W_{\text{comp}}$. The number of bits sent downwards is
\begin{equation}
	N_{\text{bits,down}}=
	\frac{K(K+1)}{2}W_{\text{comp}}+N_{\text{DL}}N_{\text{SC}}W_{\text{symbol}}.
\end{equation}
The required downwards link datarate is then
	\begin{equation}
		R_{\text{down}}=
		\frac{
			N_{\text{bits,down}}
		}
		{
			T_{\text{frame}}
		}.
	\end{equation}
However, this is only the minimum required data rate. If the data is not sent between the nodes at the same rate as it is consumed, buffers (which may incur a significant increase in die area) are needed, as discussed in Section~\ref{sec:balance}.

The reduced number of values sent from the antennas to the central unit is often used as an argument for performing distributed processing. While this is indeed the case, it must also be noted that the word lengths of the data are different. For a centralized architecture, the word length depends on the ADC, so the number of bits is proportional to $MW_{\text{ADC}}$. For a distributed architecture, $\tilde{\mathbf{y}}$ and $\mathbf{B}$ are transmitted, so the number of bits is proportional to $KW_{\text{comp}}$. Since these are sum-of-products, where one product term being the sample value, one may expect that in general $W_{\text{comp}} > W_{\text{ADC}}$. However, as $M \gg K$, the total number of bits transmitted to the central unit should still be significantly smaller. Furthermore, it is important that the intermediate values are properly scaled as more and more terms are added along the path to the central unit.

	\subsection{Balancing Computations, Communication, and Memories}\label{sec:balance}
	To obtain an optimized architecture the different types of resources must be balanced. Here, the processing, communication and memory capabilities are included.
	
\begin{figure}[tb]
		\centering
			\includegraphics[scale=0.28]{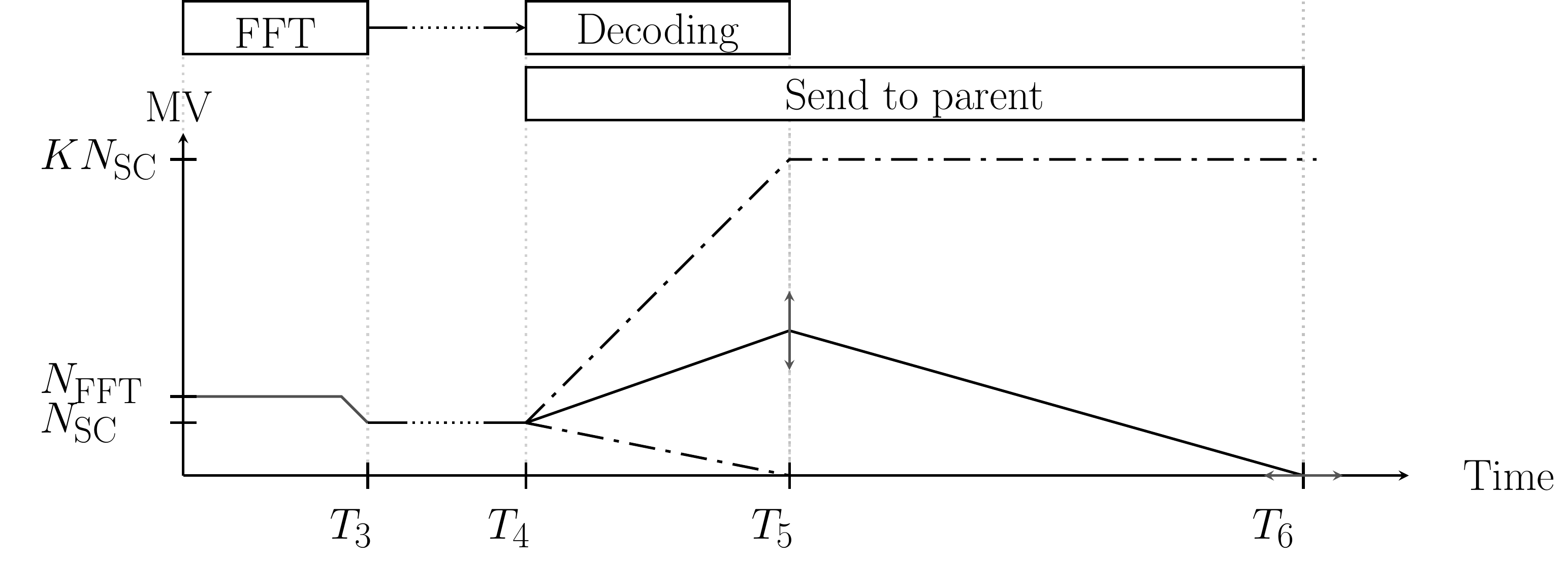}
		\caption[]{\label{fig:ul_stored_variables} Number of stored memory variables during the lifetime of one uplink OFDM symbol.}
		
	\end{figure}

	Considering the inter node communication for one uplink OFDM symbol, the number of stored variables in each node can be seen in Fig.~\ref{fig:ul_stored_variables}. From sampling the radio until the FFT is finished, the number of stored variables are the same as the number of existing variables in Fig.~\ref{fig:ul_variables}. The output data from the decoding process has no further data dependencies in the current node. These variables need to be sent to the parent node, so it can perform its own decoding process.
	
	In Fig.~\ref{fig:ul_stored_variables}, the time $T_4$ to $T_5$ is again the time taken to perform the decoding. The time $T_4$ to $T_6$ is the time taken to send the local contributions to the decoded signal vectors to the parent node. It can be noted that $T_6 \ge T_5$. When the decoding starts, the number of variables that needs to be stored locally increases due to the data expansion of the decoding, but at the same time decreases due to variables being sent to the parent node, and thus not needing to be stored. 
	
	There are two extreme cases of this behavior. The first is if $T_6$ tends to infinity. In this case, all variables must be stored locally, due to none of them being sent over the link. Clearly, this solution is not feasible. The other extreme is if $T_6 = T_5$. This means that all output variables are sent directly to the parent node and does not need to be stored locally.
	
	As described earlier, the decoding on the parent node cannot be performed until the decoding output has been sent over the link. The implication of this is that the system should be designed such that the rate of processing and the rate of sending variables between nodes are the same.
	
	The requirements on the downwards communication however are not as strict. The data that is propagated from the CCU to the nodes in the tree is not processed on the way downwards, but rather just forwarded to the next node. This has the implication that it is not required to send the data in the same rate as it is processed. It does however prevent the need for large buffers in each node, making it desirable. Feeding the nodes with data is a rather straightforward trade-off between link data rate and buffer size.

\section{Scheduling}\label{sec:scheduling}
	The computational tasks and data dependencies when using ZF processing can be seen in Fig.~\ref{fig:buffer_size}. This schedule is not correctly scaled, but rather made to illustrate the data dependencies and the need for a better realization. It is, e.g., clear that there is time at the end of the frame where no operations are currently performed. Therefore, it makes sense to move parts of the computations there to obtain a better utilization of the processing elements. This will come at a cost of memory as the data must be stored rather than processed directly.
	
\begin{figure}[tb]
		\centering
			\includegraphics[scale=0.53]{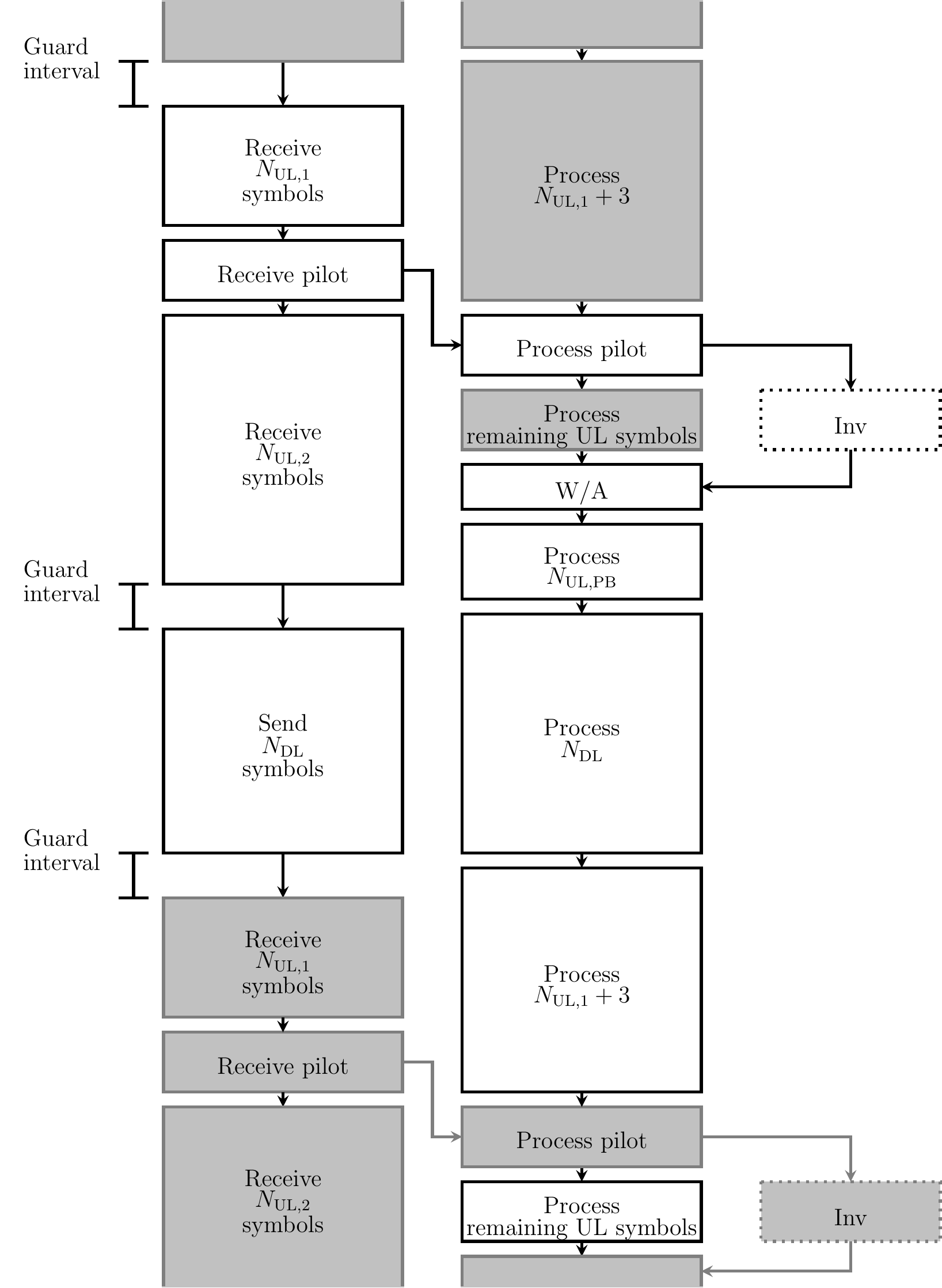}
		\caption[]{\label{fig:flow} Schedule of the computational tasks in the asymptotic case. The white blocks corresponds to one frame.}
	\end{figure}

	Here, a node with only one processing element is considered. The processing element is assumed to support the required number of operations per sample for the asymptotic case, i.e., $\hat{N}_{\text{OPS}} \geq N_{\text{OPS,asymptotic}}$. The schedule is created as in Fig.~\ref{fig:flow}.
	Initially, the node will wait for the pilot OFDM symbol. The computations for determining the precoding/decoding vector is then started. This includes an FFT, performing the channel estimation and computing the $\mathbf{B}_i$ matrix. When the inverted matrix is received, the precoding/decoding vector is computed. After this stage, the uplink and downlink symbols can be processed. In order to reduce the number of uplink symbols that needs to be buffered before processing, $N_{\text{UL,PB}}$ uplink symbols is processed. All downlink symbols is then processed in order to meet their deadlines. When the downlink symbols are finished, the node computes $N_{\text{UL,1}}+3$ uplink symbols is processed. Two uplink symbols can be processed when the last downlink symbol is transmitted and during the guard interval. Another uplink symbol can be computed when pilot of the next frame is sampled. The remaining $N_{\text{UL}}-N_{\text{UL,PB}}-N_{\text{UL,1}}-3$ uplink symbols are processed when the node waits for the inverted matrix for the next frame.
	
	As can be seen, the processing is fully deterministic for the asymptotic case, and, hence, a simple control unit can be implemented, where the different system parameters can be configured. For the non-asymptotic case, the same general structure is implemented. However, as the processing is possibly distributed differently within the frame, a slightly more flexible control unit is required. Alternatively, the control signals can be stored in a RAM acting as an instruction memory.
	
	\subsection{System Level Scheduling\label{sec:sls}}
	For the computational tasks $\tilde{\mathbf{y}}_i$ and $\mathbf{B}_i$ in Table~\ref{tab:num_op} there are inter-node data dependencies, as described in Section~\ref{sec:mapping}. Before the local PE operation can be performed, the corresponding contributions from the child nodes must be sent over the inter-node link. The latency of sending a value over the link is $T_{\text{link}}$. For each level in the tree, the $\tilde{\mathbf{y}}_i$ and $\mathbf{B}_i$ computations must be skewed by this amount in order for the parent node to receive the data before processing.	

\section{Architecture}
	In this section, an architecture for the node is proposed. The main components in the architecture are the off-chip I/O, processing core, memory system and the RF-chain.
	
\begin{figure}[tb]
		\centering
			\includegraphics[scale=0.48]{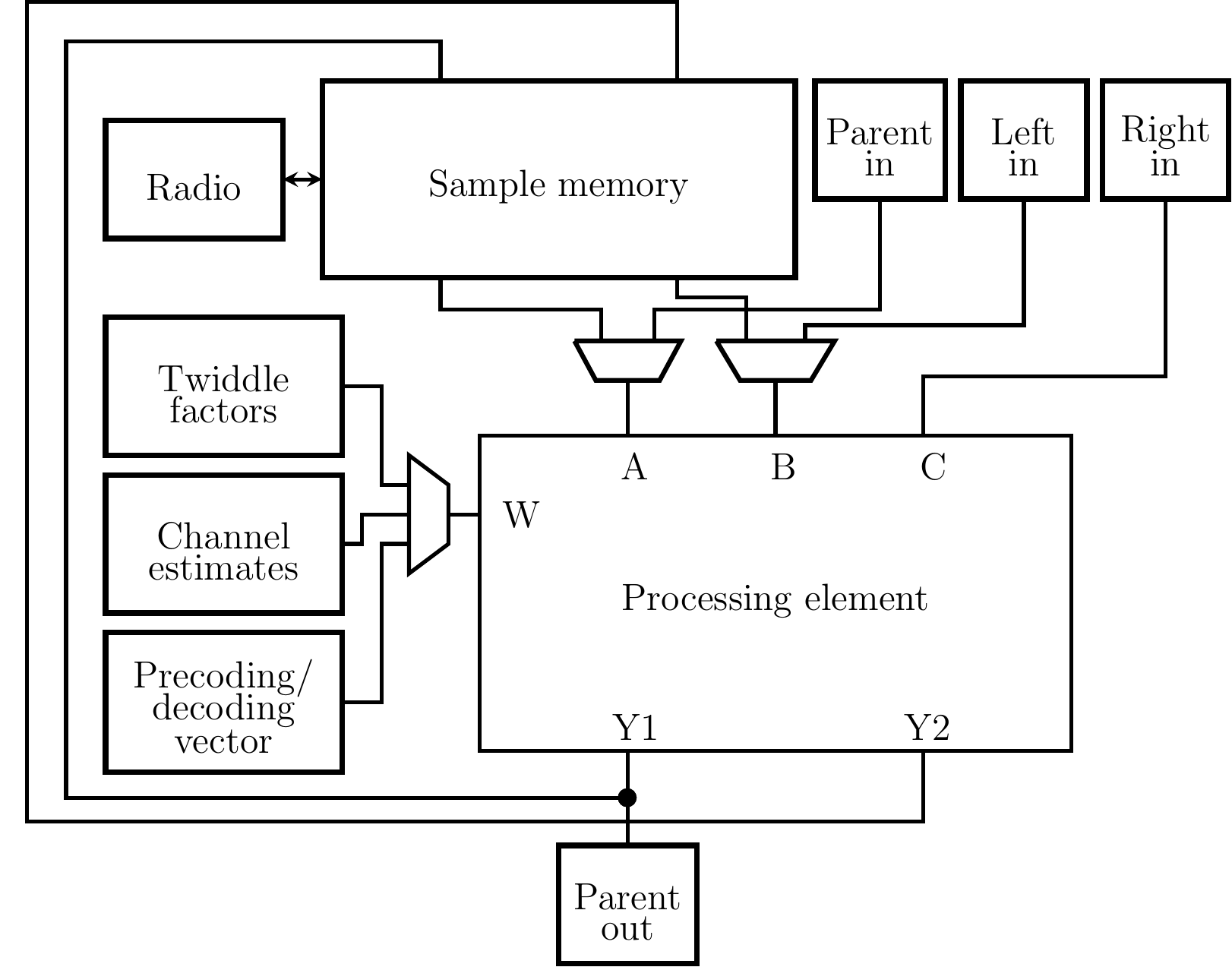}
		\caption[]{\label{fig:node_arch} Proposed node architecture (not all connections shown).}
		
	\end{figure}

	
	As seen later on in Section~\ref{sec:dse} many system scenarios can be covered with a single processing element in each node. Hence, we focus on that here.
	Further inspection of the arithmetic operations in Fig.~\ref{fig:arith_op} reveals that each input port of the processing element is connected only to a few specific data. This means that not all types of data need to be fed into any port of the PE. For instance, only the twiddle factors, channel estimates or the precoding/decoding vector is connected to one input of the multiplier. Taking this into account leads to the proposed node architecture shown in Fig.~\ref{fig:node_arch}. The node architecture uses a processing element as shown in Fig.~\ref{fig:selected_pe}.

\begin{figure}[tb]
		\centering
				\includegraphics[scale=0.5]{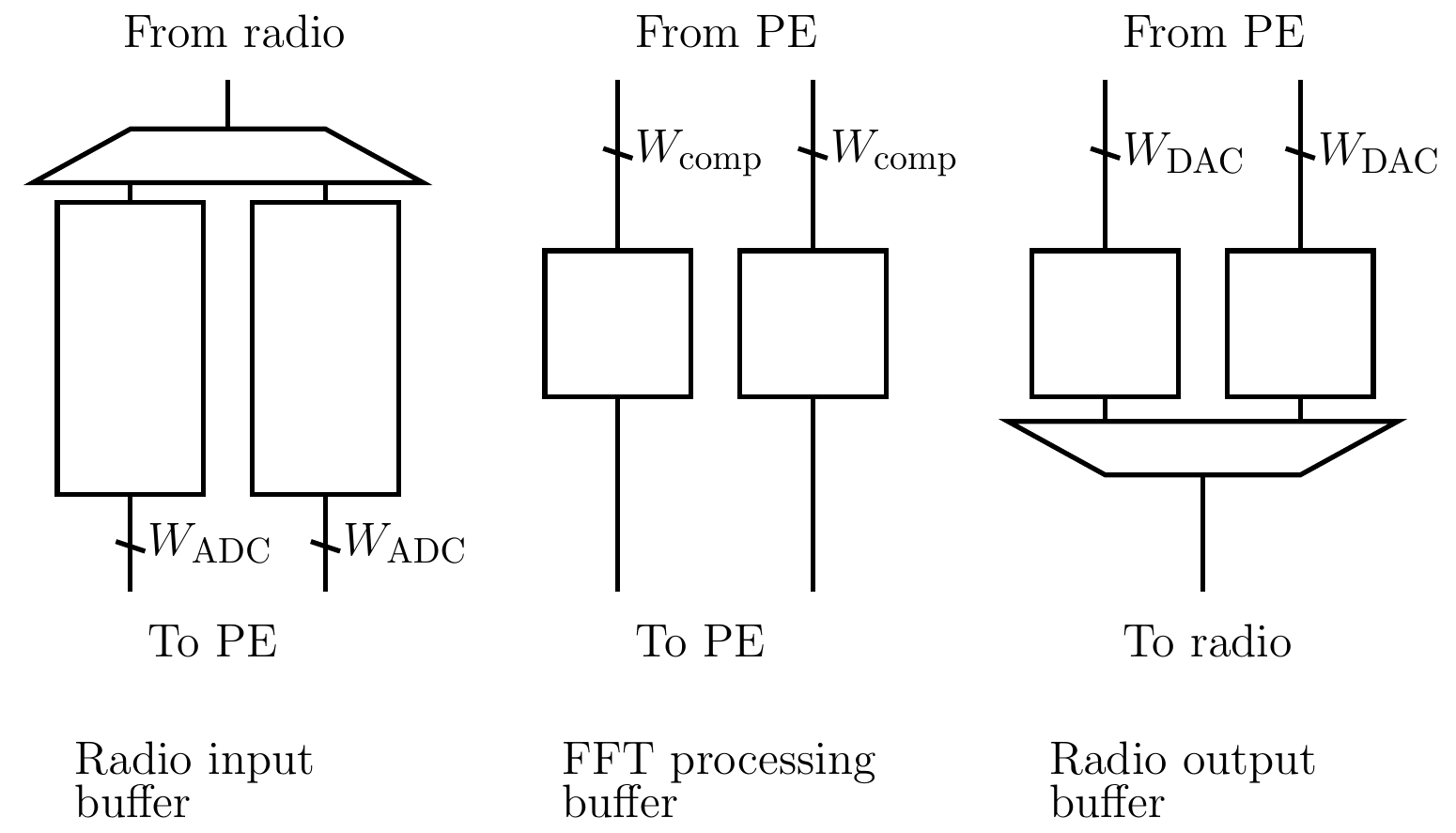}
		\caption[]{\label{fig:mimo_memory} Structure of the sample memory in Fig.~\ref{fig:node_arch}. }
		
	\end{figure}

	The twiddle factor memory can be implemented as a ROM, since the twiddle factors are static. The channel estimates and precoding/decoding vector memories are single port memories, that can either be written or read in one cycle. Although during precoding and decoding, only the precoding/decoding vector is required, the channel estimates must be stored until all precoding/decoding values are computed, and, hence, both must be stored. For simplicity, we select to have a separate memory allocation for the channel estimate, instead of using e.g. the sample memory.
	The sample memory is more complex and is divided into three separate memories as shown in Fig.~\ref{fig:mimo_memory}.
The first memory is the radio input buffer which stores raw data from the AD converter. The size of this memory is
	\begin{equation}
		\text{Mem}_{\text{input}}=
		N_{\text{UL,buffered}}N_{\text{FFT}}W_{\text{ADC}} \ \text{bits}.
	\end{equation}
	
	The FFT processing buffer is used when performing the FFT/IFFT computations and its size is
	\begin{equation}
		\text{Mem}_{\text{processing}}=
		N_{\text{FFT}}W_{\text{comp}} \ \text{bits}.
	\end{equation}
	
	The last memory is the radio output buffer which holds the finished downlink OFDM symbols that are to be sent to the DA converter. The cyclic prefix of the OFDM symbol is also fetched from the memory. Its size is
	\begin{equation}
		\text{Mem}_{\text{output}}=
		N_{\text{FFT}}W_{\text{DAC}} \ \text{bits}.
	\end{equation}
	In Fig.~\ref{fig:mimo_memory}, the memories are shown as two-port memories being able to read and write simultaneously. In many cases, it may be beneficial to use two single-port memories of the half the size instead. For the input and output buffers, it is straightforward to use memories alternating reading and writing. For the FFT processing buffer, it is also possible using e.g. the approach in \cite{Ma2000}.

	All memory sizes and word lengths for the architecture in Fig.~\ref{fig:node_arch} are summarized in Table~\ref{tab:memory}.
	The exact required word lengths should be based on system level simulations, which is left for future work.	
		\begin{table}
			\centering
			\caption{Memory Sizes for a Node.}
			\label{tab:memory}
				\begin{tabular}{  l | c c  }
				\hline
				\textbf{Memory} & \textbf{\#Words} & \textbf{Word length}\\ \hline
				Input buffer & $N_{\text{UL,buffered}}N_{\text{FFT}}$ & $W_{\text{ADC}}$ \\
				FFT processing buffer & $N_{\text{FFT}}$ & $W_{\text{comp}}$ \\
				Output buffer & $N_{\text{FFT}}$ & $W_{\text{DAC}}$ \\
				Channel estimates & $K$ & $W_{\text{comp}}$ \\
				Precoding/decoding vector & $K$ & $W_{\text{comp}}$ \\
				Twiddle factors (ROM) & $N_{\text{FFT}}/2$ & $W_{\text{TF}}$ \\
				\hline
			\end{tabular}
			
		\end{table}
		
\section{Example: LTE-Like System Specifications}\label{sec:lte_example}
	Here, the requirements for an LTE-like system using ZF processing are considered. This is the typical specification considered in most earlier work. The system specifications can be seen in Table~\ref{tab:spec_lte}.
		For this specification, the throughput is 
	\begin{equation}
		\frac{N_{\text{SC}}N_{\text{DL/UL}}KW_{\text{symbol}}}{T_{\text{frame}}} = 384\ \text{Mb/s}
	\end{equation}
	in each direction. The centralized matrix inversion can be performed either using an exact \cite{Ingemarsson2016} or an approximate \cite{Wu2014,Liang2015,Abbas2016} algorithm. As shown in \cite{Gustafsson2017}, the complexity is similar for the best exact algorithm and a Neumann series approximation with three terms. In both cases a $20 \times 20$ matrix inversion can be performed in less than $40 \ \mu$s using one processing element running at $200$ MHz.
	
	\begin{table}
		\centering
		\caption{Specifications of the LTE-Like System in the Example.}
		\label{tab:spec_lte}
			\begin{tabular}{  l | r || l | r@{ }l }
			\hline
			\textbf{Name} & \textbf{Value} & \textbf{Name} & \multicolumn{2}{c}{\textbf{Value}} \\ \hline
			$K$ & $20$& $T_{\text{frame}}$ & $0.5$&ms \\
			$N_{\text{FFT}}$ & $2048$ & 	$T_{\text{inv}}$ & $40$& $\mu$s \\ 
			$N_{\text{SC}}$ & $1200$& $W_{\text{comp}}$ & $12+12$&bits \\ 
			$N_{\text{UL,1}}$ & 0  & $W_{\text{symbol}}$ & $2+2$ & bits \\
			$N_{\text{UL,2}}$ & 2 & $W_{\text{ADC}}$ & $6+6$ &bits\\ 
			$N_{\text{DL}}$ & 2  & $W_{\text{DAC}}$ & $6+6$ & bits\\ 
			$T_{\text{link}}$ & $0.5\mu$s  & $f_{\text{sample}}$ & $30.72$ &MHz\\
			\hline
		\end{tabular}
	\end{table}
	
	
	Based on these specifications and (\ref{eq:opsavg}) and (\ref{eq:opscritical}), we get
	\begin{equation}
	N_{\text{OPS,avg}}=9.95
	\end{equation}
	and
	\begin{equation}
	N_{\text{OPS,critical}}=
	\max\left(9.23, \ 11.28\right)=11.28.
	\end{equation}
	In this case, $T_{\text{inv,A}}=8.27 \ \mu$s and $T_{\text{inv,B}}=110 \ \mu$s. Therefore, based on (\ref{eq:nops}) and (\ref{eq:pesfclk}), selecting one PE running at $12f_{\text{sample}} = 368.64$ MHz is sufficient, leading to $\hat{N}_{\text{OPS}}=12$. In this case, the second downlink OFDM symbol imposes the limit on the number of computational resources. Hence,  $N_{\text{UL,PB}}=0$ and $N_{\text{UL,buffered}}=2$.

\begin{figure}[tb]
		\centering
				\includegraphics[scale=0.5]{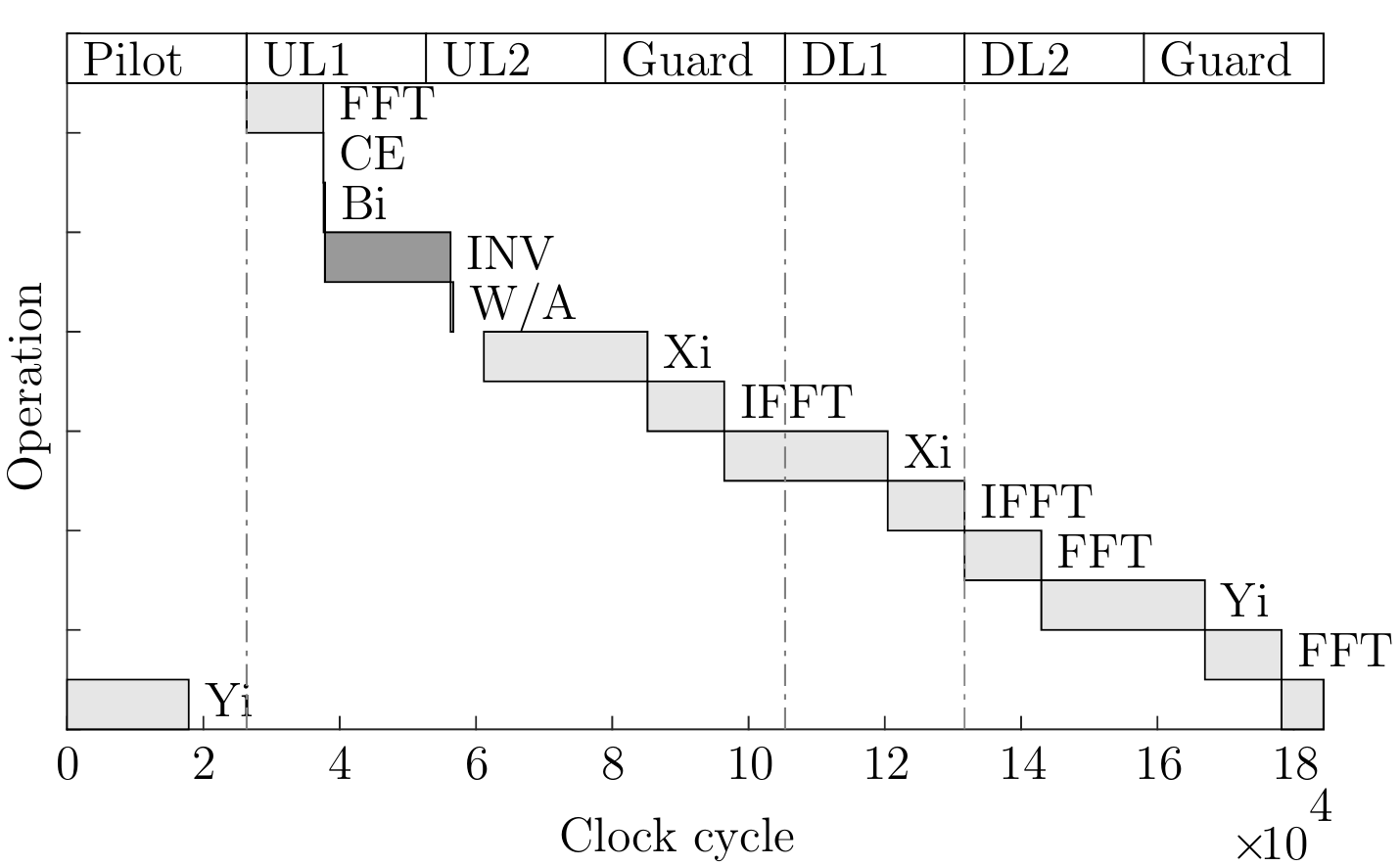}
		\caption[]{\label{fig:sched_real} Schedule for the LTE-like system with ZF processing.}
	\end{figure}

	 A schedule for the computations in the LTE-like system is derived and can be seen in Fig.~\ref{fig:sched_real}. Deriving the schedule is rather straightforward since all tasks are performed sequentially. The slack time between determining the local precoding/decoding vector, $\mathbf{W}_i$/$\mathbf{A}_i$, and the start of the precoding, $x_i$, can be utilized to modify the specification, as discussed below.

	The size of the radio input buffer is
	\begin{equation}
		\text{Mem}_{\text{input}}=
		N_{\text{UL,buffered}}N_{\text{FFT}}W_{\text{ADC}}=
		48 \ \text{kb,}
	\end{equation}
	the size of the FFT processing buffer is
	\begin{equation}
		\text{Mem}_{\text{processing}}=
		N_{\text{FFT}}W_{\text{comp}}=
		48 \ \text{kb,}
	\end{equation}
	and the size of the radio output buffer is
	\begin{equation}
		\text{Mem}_{\text{output}}=
		N_{\text{FFT}}W_{\text{DAC}}=
		24 \ \text{kb.}
	\end{equation}
	Hence, a total of 120 kb of memory is required in each node. In addition, 960 more bits are needed for the channel estimates and precoding/decoding vector.
	
		
		As was shown in Section \ref{sec:resource_allocation}, the computations and communication needs to be performed at the same rate. With the selected number of operations performed per sample, $N_{\text{OPS}}$, the data rates are
		\begin{equation}
			R_{\text{up}}=
			\hat{N}_{\text{OPS}}f_{\text{sample}}W_{\text{comp}}\approx
			8.847 \ \text{Gbps}
		\end{equation}
		and
		\begin{equation}
			R_{\text{down}}=
			\hat{N}_{\text{OPS}}f_{\text{sample}}W_{\text{symbol}}\approx
			1.475 \ \text{Gbps}.
		\end{equation}
	
		The available slack time can be used to modify the specifications of the system. By tweaking the parameters and redoing the calculations, we can investigate which configurations are supported with $\hat{N}_{\text{OPS}}=12$.
		For example, the matrix inversion time can be increased up to $54.2 \ \mu$s, with exactly the same node architecture. Alternatively, the number of users can be increased to $K=21$, assuming that the matrix inverse time increases cubically. For $K=30$ and the same assumption, $\hat{N}_{\text{OPS}}$ can be selected to $28$. In this case, either one PE running at $860.16$~MHz or two PEs running at $430.08$~MHz can be used\footnote{Naturally, any combination of \NPE~and \FCLK~is valid as long as (\ref{eq:pesfclk}) holds. However, if multiple PEs are used, the memory architecture may need to be modified.}. Alternatively, for the example, $N_{\text{hops}}$ can be increased up to $22$, leading to a maximum of $M=2^{22}-1$ antennas, assuming a binary tree. Even though this can be further increased by increasing $\hat{N}_{\text{OPS}}$, this should not pose a limitation in most cases.
	
		If we want to process one uplink symbol before the first downlink symbol, i.e., $N_{\text{UL,PB}}=1$, we must select $\hat{N}_{\text{OPS}}=17$. To move the pilot symbol one symbol closer to the downlink symbols, i.e., $N_{\text{UL,1}}=1$ and $N_{\text{UL,2}}=1$, again $\hat{N}_{\text{OPS}}=17$, although this equality does not hold in general. Halving the matrix inversion time leads to $\hat{N}_{\text{OPS}}=15$ in both cases. This illustrates that when the critical paths are limiting, increasing the computational capabilities in the CCU, i.e., decreasing the matrix inversion time, leads to reduced computational requirements in the nodes.
		
		Naturally, any valid combination of these modifications can be realized.
		
\section{Design Space Exploration}\label{sec:dse}
\begin{figure}[tb]
		\centering
			\includegraphics[scale=0.5]{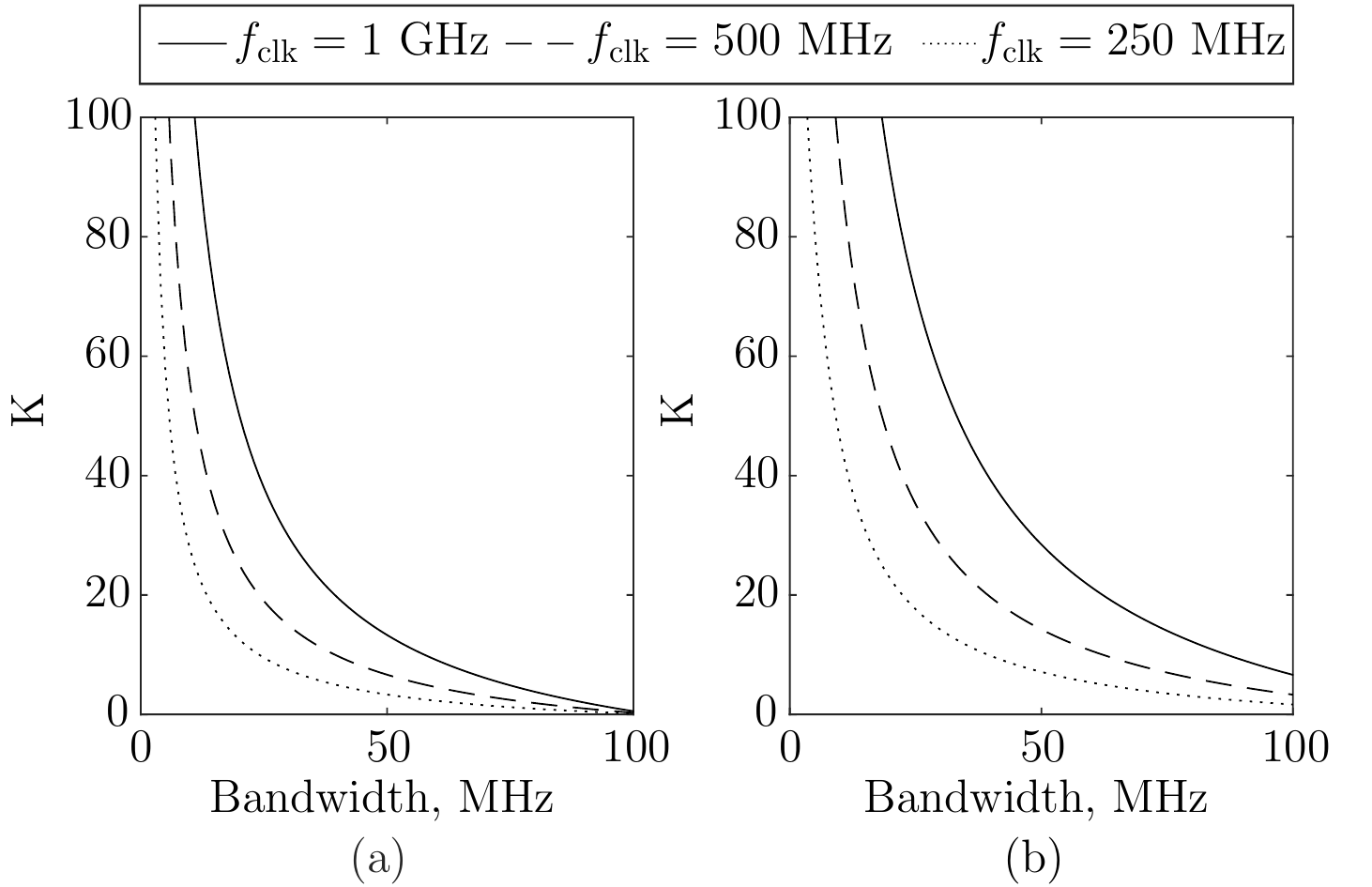}
		\caption[]{\label{fig:exploration} Bound on bandwidth and terminals using a single processing element at a given $f_{\text{clk}}$ for the LTE-like case. Configurations with (a) large and (b) small number of OFDM symbols.}
	\end{figure}

	The clock frequency required in the LTE-like example, is not a problem to achieve in a modern process technology through, e.g., pipelining, which is straightforward since the execution is deterministic. Hence, it is possible to change the bandwidth and/or the number of terminals. Here, we consider three different clock frequencies up to $1$~GHz for a system otherwise as in the LTE-like case. Figure~\ref{fig:exploration} shows the bounds on bandwidth and number of terminals for a given clock frequency. In Fig.~\ref{fig:exploration}(a) the asymptotic case is shown, $N_{\text{OPS}}=N_{\text{OPS,asymptotic}}$, meaning that the number of OFDM symbols in each frame is large. In Fig.~\ref{fig:exploration}(b) the frame format in the LTE-case is used. In both cases the average number of computations over an entire frame is used. Thus, it is assumed that the matrix inversion is performed fast enough to not influence the required number of operations per second, i.e., $T_{\text{inv}}\leq T_{\text{inv,A}}$.
	
	It is noted from Fig.~\ref{fig:exploration} that increasing the channel bandwidth by a factor two, roughly requires that the number of simultaneous terminals are reduced by the same factor.
	
	Here, the length of the FFT, $N_{\text{FFT}}$, and the number of subcarriers utilized, $N_{\text{SC}}$, is scaled linearly with the bandwidth of the channel. This is usually not the case, since the FFT length is favorably selected as a power of two. The plots still give a good estimate of the available design space.
\section{Conclusions}
	In this work, a scalable system architecture using distributed processing was proposed for the base station in a massive MIMO system. It was shown that the computations associated with each antenna can be distributed and in most of the earlier studied use cases only a simple single processing element running at a few hundred MHz and a modest amount of memory are required. It was further shown that it is feasible to have a simple synchronous control of the nodes and that the inter-node communication can be handled by one or a few high-speed serial links. All computations required by adding an antenna are handled by the introduced additional node.
	
	The case of connecting the nodes as a binary tree was primarily studied, although the architecture is readily extended to a $K$-ary tree. It is here worth noting that an array architecture with static scheduling will behave as a binary or ternary tree, and, hence, the same concept can be used for an array interconnect with additional simple routing logic surrounding the processing node. As the processing core is so small, it is also of interest to possibly have more than one node in a chip, reducing the amount of inter-chip communication channels. 
	The exact granularity is left for future work.

	The architecture supports conjugate beamforming, zero forcing, and MMSE processing. In the latter two cases, a matrix inversion is performed in a central control unit, but all other computations are distributed. The impact of the matrix inversion latency and pilot position on the computational requirements in the node are studied and related.


%

\ifCLASSOPTIONcaptionsoff
  \newpage
\fi



\bibliographystyle{IEEEtran}
\bibliography{IEEEabrv,ES_Conf_abrv,mimo}

\end{document}